\documentclass[11pt,a4paper]{article}

\usepackage{cite}
\usepackage[cmex10]{amsmath}
\usepackage{amsmath}
\usepackage{amssymb}
\usepackage{booktabs}
\usepackage{color}
\usepackage{times}
\usepackage{paralist}
\usepackage{subfigure}
\usepackage{float}
\usepackage{epstopdf}
\usepackage{authblk}
\usepackage{graphicx}

\renewcommand{\u}{\mathbf{u}}
\newcommand{\x}{\mathbf{x}}
\renewcommand{\r}{\mathbf{r}}
\newcommand{\w}{\mathbf{w}}
\renewcommand{\v}{\mathbf{v}}

\newcommand{\y}{\mathbf{y}}
\renewcommand{\b}{\mathbf{b}}
\renewcommand{\c}{\mathbf{c}}
\renewcommand{\k}{\mathbf{k}}

\newcommand{\A}{\mathbf{A}}
\newcommand{\B}{\mathbf{B}}
\newcommand{\E}{\mathbf{E}}
\newcommand{\M}{\mathbf{M}}

\newcommand{\Q}{\mathbf{Q}}
\newcommand{\R}{\mathbf{R}}
\newcommand{\Ss}{\mathbf{S}}
\newcommand{\G}{\mathbf{G}}
\newcommand{\F}{\mathbf{F}}
\newcommand{\D}{\mathbf{D}}
\newcommand{\C}{\mathbf{C}}
\newcommand{\I}{\mathbf{I}}
\newcommand{\Hb}{\mathbf{H}}
\newcommand{\Ib}{\mathbf{I}}
\newcommand{\X}{\mathbf{X}}

\let\mathnumsetfont\mathbb
\newcommand\Rset{\mathnumsetfont R}       
\newcommand\Wset{\mathnumsetfont W}       

\begin{document}
\title{Robust Energy-Constrained Frequency Reserves\\from Aggregations of Commercial Buildings}
\author[1]{Evangelos~Vrettos\thanks{vrettos@eeh.ee.ethz.ch}}
\author[2]{Frauke~Oldewurtel\thanks{oldewurtel@berkeley.edu}}
\author[1]{G\"{o}ran~Andersson\thanks{andersson@eeh.ee.ethz.ch}}
\affil[1]{EEH-Power Systems Laboratory, Swiss Federal Institute of Technology (ETH) Zurich, Switzerland}
\affil[2]{Department of Electrical Engineering \& Computer Sciences, University of California, Berkeley, USA}

\maketitle

\begin{abstract}
It has been shown that the heating, ventilation, and air conditioning (HVAC) systems of commercial buildings can offer ancillary services to power systems without loss of comfort. In this paper, we propose a new control framework for reliable scheduling and provision of frequency reserves by aggregations of commercial buildings. The framework incorporates energy-constrained frequency signals, which are adopted by several transmission system operators for loads and storage devices. We use a hierarchical approach with three levels: (i) reserve capacities are allocated among buildings (e.g., on a daily basis) using techniques from robust optimization, (ii) a robust model predictive controller optimizes the HVAC system consumption typically every $30$ minutes, and (iii) a feedback controller adjusts the consumption to provide reserves in real time. We demonstrate how the framework can be used to estimate the reserve capacities in simulations with typical Swiss office buildings and different reserve product characteristics. Our results show that an aggregation of approximately $100$ buildings suffices to meet the $5$~MW minimum bid size of the Swiss reserve market.
\end{abstract}

\section{Introduction}
In many countries, large amounts of renewable energy sources (RES) are integrated in the power system and they increase the operational uncertainty due to their fluctuating nature. In a power system, supply and demand of electric power must be balanced to keep frequency and voltage close to their nominal values. To maintain this balance, transmission system operators (TSOs) procure ancillary services (AS), e.g., frequency and voltage control, traditionally from conventional generators. With increasing RES shares, the need for AS is also increasing \cite{makarov_operational_2009}. For this reason, there has been a recent interest in offering AS by flexible loads, which is known as demand response (DR). If properly aggregated, loads can provide AS without environmental impacts and, possibly, more efficiently and at a lower cost compared to generators \cite{callaway_achieving_2011}.

Loads with thermal storage are suitable for DR because their consumption can be shifted in time without impact on consumer comfort. A lot of research focused on modeling, control, and estimation of large aggregations of thermostatically controlled loads (TCLs), i.e., small residential loads such as refrigerators, air conditioners, space and water heaters \cite{kalsi_development_2012,Vrettos2012,Perfumo2012,mathieu_state_2013,VrettosPSCC2014}. The heating, ventilation, and air conditioning (HVAC) systems of commercial buildings are also interesting for DR because they are large consumers with high thermal inertia, and they usually include a building automation and communication system that facilitates DR implementation \cite{IREP2013}. Recently, DR applications with commercial buildings has been the topic of several papers some of which are summarized below. The role of information systems and automated model-based control for energy efficiency and DR was discussed in \cite{pietteLBNL2012}. Commercial HVAC systems are typically complex with many control variables and cascaded loops, and thus low-order models are used for control purposes \cite{Sturzenegger2013Experiment,Lehmann_2013,Goddard2014}. Research on DR with commercial buildings has focused mainly on ``slow'' services. For example, \cite{Braun1990,Oldewurtel2010,Ma2012,VrettosECC2013} proposed optimization-based approaches and model predictive control (MPC) for peak shedding or load shifting to minimize energy costs.

This paper is concerned with frequency control with commercial buildings, in particular secondary frequency control (SFC), which is also known as automatic generation control (AGC), load frequency control (LFC), automatic frequency restoration reserve (FRR), or regulation service. Recently, some authors have considered either scheduling or provision of ``fast'' AS from commercial buildings including SFC reserves. Ref. \cite{kiliccote_field_2012} experimentally investigated the feasibility of offering up and down regulation products with university campus buildings, and identified baseline prediction and latency as potential obstacles for implementation. The authors of \cite{Hao2013Allerton,Hao2013ACC} investigated SFC via fan power control in buildings with variable air volume (VAV) systems. Based on simulations, the authors found that up to $15\%$ of a building's fan power can be offered as reserves without significant loss of comfort, if the SFC signal is within the frequency band $f \in [1/(10~\textrm{min}), 1/(4~\textrm{sec})]$. In \cite{LinTSG2014}, a control approach similar to that of \cite{Hao2013Allerton,Hao2013ACC} was experimentally validated in a real building, but without considering frequencies below $1/(10~\textrm{min})$ to avoid effects on chiller power consumption. The follow-up work \cite{LinIEEESGCom2013} included chiller control, which enlarged the frequency band of SFC signals to $1/(60~\textrm{min})$. Ref. \cite{Kim2014} investigated SFC by direct control of a heat pump's (HP) compressor power using a variable speed drive motor in a water-based HVAC system.

Apart from accurate tracking of the SFC signal, a TSO needs guarantees that the reserve capacity of commercial buildings will be \emph{reliably} available. Because the buildings are energy-constrained resources reserve scheduling is required, in particular if the SFC signal is not approximately zero-energy over short periods but can be biased towards one direction over long periods of time. A priori reserve scheduling allows buildings to participate in markets for such reserve products without compromising occupants' comfort. Ref. \cite{MaasoumyACC2014paper1} addressed this issue by developing an MPC-based method to quantify the flexibility of a commercial building and a contractual framework to declare it to the utility. In this paper, we follow this line of research and consider robust reserve scheduling for \emph{aggregations} of commercial buildings.

The main contributions of this paper are threefold. First, we propose a new framework to estimate the SFC reserve \emph{capacity} that can be reliably offered by an aggregation of commercial buildings considering weather conditions, occupancy, electricity prices, reserve payments, and comfort zone. The framework builds on a hierarchical control scheme with three levels, namely reserve scheduling and allocation, HVAC control, and reserve provision, and is based on robust optimization and MPC. The framework actively allocates reserves among aggregation's buildings based on their individual characteristics, which is expected to maximize the reserve potential in markets with typical requirements such as constant reserve capacity over a minimum duration and/or equal up- and down-reserve capacities. This is in contrast to \cite{Hao2013Allerton,Hao2013ACC,LinIEEESGCom2013} that estimated the capacity of a group of buildings by simply scaling up the estimated capacity of a single building. Second, we propose new methods to estimate reserve capacities in case of energy-constrained frequency signals, a practice adopted by several TSOs for loads or storage devices, e.g., \cite{PJMASmanual}. This is a significant advantage of our method compared to \cite{MaasoumyACC2014paper1}. A relevant approach is taken in \cite{Barooah2015HICSS}, where the reliable AS capacity from a commercial building is expressed as a function of the AS signal's frequency. In contrast to \cite{Barooah2015HICSS}, we are interested in the dependence of capacity on the integral of the SFC signal, i.e., on the signal's bias directly, which we believe is the major limiting factor when extracting reserves from energy-constrained resources. Third, we derive upper bounds on the untapped potential for SFC by different building types and for different reserve product characteristics such as duration, symmetry, and energy content.

Although in some cases the SFC signal is approximately zero-energy, this is not generally true for all power systems. Ideally the bias of the SFC signal would be obtained from generators, e.g., through activation of tertiary control reserves or in a nearly real-time (e.g., 5 minute) market. However, nearly real-time markets do not exist in many control areas, and activation of tertiary control reserves is not always more economical than SFC reserves, because it highly depends on the the availability of cheap and flexible generators. In addition, with increasing RES penetration in power systems, the share of conventional generators in the production mix will decrease. Despite the improvements in forecasting techniques, the RES forecasts will never be perfect and biased forecast errors will translate into biased SFC signals. With fewer controllable generators in the production mix, other resources will have to absorb these biases. For this reason, we believe that accounting for biases in the SFC signal will be very important in the future, possibly also in power systems where the SFC signal is today approximately zero-energy. To the authors' best knowledge, this is the first time that methods are proposed to allow building aggregations to systematically schedule the reserve capacity that they can offer depending on how much bias exists in the SFC signal.

This paper includes substantial extensions over our previous work \cite{VrettosIFAC2014}, where some preliminary results were presented. A major improvement is related to modeling of the SFC reserve requests within the reserve scheduling and building control optimization problems. This paper considers also energy-constrained SFC signals apart from conventional unconstrained signals, whereas \cite{VrettosIFAC2014} considered only the latter. In the absence of energy constraints, the worst case reserve request is equal to either the up- or down-reserve capacity along the whole scheduling horizon. The reserve scheduling problem for this case was cast as a robust optimization problem with additive uncertainty in \cite{VrettosIFAC2014}. On the contrary, the determination of the worst case reserve request along the scheduling horizon is more involved with energy constraints due to time coupling. In this paper, we model an energy-constrained SFC signal as an uncertain variable that lives in a polyhedral uncertainty set and enters multiplicatively in the constraints. This allows us to reformulate the robust reserve scheduling problem into a deterministic tractable one. Another main extension is related to the number and type of actuators that provide reserves in each building. In contrast to \cite{VrettosIFAC2014} that assumed either only heating or only cooling actuators per building, the proposed formulations can be used to allocate reserves among different types of actuators within the same building. Therefore, the proposed formulations are more general than the ones in \cite{VrettosIFAC2014}. In addition, the simulation studies of this paper contain new material compared with \cite{VrettosIFAC2014}. We analyze recorded data of the Swiss SFC signal to illustrate that the SFC signal is not necessarily approximately zero-energy, but it can be significantly biased over periods of several hours. The performance of the control framework is then evaluated using an energy-constrained SFC signal, whereas the sensitivity analysis is performed considering signals with and without energy constraints.

The rest of the paper is organized as follows: Section~\ref{FreqCtrl} summarizes some important aspects of power system SFC, Sections~\ref{Modeling} and~\ref{ControllerDesign} introduce the modeling and control framework, the performance of the framework is demonstrated and evaluated in Section~\ref{ControllerPerformance} in a simulation example, a sensitivity analysis is performed in Section~\ref{SensitivityAnalysis}, and Section~\ref{Conclusion} concludes the paper.

\section{Frequency Control in Power Systems} \label{FreqCtrl}
Typically, a TSO controls frequency in three steps: primary, secondary and tertiary control. Primary control is a distributed, proportional controller that stabilizes the frequency after a disturbance. Secondary control is a centralized, proportional-integral (PI) controller that restores the frequency to its nominal value and maintains the desired exchanges between neighboring control areas. Tertiary control releases secondary control in case of large disturbances and is typically manually activated.

Before continuing, we summarize some important aspects of scheduling and activation of SFC in power systems \cite{ENTSOE_AS_survey2014}, which will be particularly relevant to the sensitivity analysis of Section~\ref{SensitivityAnalysis}. In most of Europe, SFC reserves are procured in a market setting, i.e., the generators bid their reserve capacity and price in weekly or daily auctions. The requirements of these auctions and the characteristics of the SFC reserve products vary from country to country. The minimum bid size is typically in the range $[1,10]$ MW, e.g., $5$~MW in Switzerland \cite{swissgridAS}. In many countries, only symmetric reserves, i.e., equal up- and down-reserve capacities, are allowed, whereas in other countries asymmetric reserves are also accepted. The reserve energy is requested from the generators via a signal sent by the TSO, typically every $2-4$ seconds. There are two main activation rules: (a) the pro-rata activation, where the reserve energy is proportional to the capacity, and (b) the merit-order activation, where the reserves are requested based on the short term marginal costs of generators. The remuneration of reserve energy is also country dependent but it is typically separate from the reserve capacity remuneration. In some countries, e.g., in Switzerland, the reserve energy remuneration is coupled with the energy price in the spot market. As a final remark, note that we treat the SFC signal as uncertain in this paper because it is unknown at the time when the reserve capacities are determined.

\section{Modeling} \label{Modeling}
\subsection{Building Aggregation} \label{BldModeling}
We consider buildings with integrated room automation (IRA) systems, where heating, cooling, ventilation, blinds, and lighting are jointly controlled. IRA is typically used in office buildings because it provides high comfort while being energy efficient \cite{Oldewurtel2011}. We represent building \emph{thermal} dynamics using a well-tested $12^{\textrm{th}}$ order multiple-input-multiple-output bilinear model \cite{Lehmann_2013}. This model is based on the widespread thermal resistance-capacitance network approach, and was validated against the well-known building simulation software TRNSYS \cite{Lehmann_2013}. The model's satisfactory accuracy and its relatively low complexity make it suitable for MPC. A model similar to the one considered here was used in an MPC implementation in a real building, and it was found that it captures the building thermal dynamics well \cite{Sturzenegger2013Experiment}. Furthermore, this model is flexibly customizable and allows us to perform large-scale simulation studies with different representative building types.

The original model is bilinear between inputs $u$ and disturbances $v$ (e.g., blind position and solar radiation), as well as between inputs and states $x$ (e.g., blind position and room temperature). If the disturbances are fixed, e.g., according to their predicted values, the bilinearities between $u$ and $v$ vanish and the system becomes time-varying. However, the bilinearities between $x$ and $u$ remain. For optimization purposes, sequential linear programming (SLP) can be applied by iteratively linearizing the bilinear $x$ terms around the most recently calculated $x$ trajectory and solving the resulting linear program (LP) until convergence \cite{Oldewurtel2011}.

If SLP is applied, the dynamics of a building $b$ can be described by the linear time-varying model
\begin{align}
x^b_{t+1} &= A^bx_t^b +B^b_t u^b_t +E^bv^b_t +R^b\Delta{u}^b_t~, \label{OriginalModel1}
\end{align}
where $x^b_t \in \Rset^{n_x}$ denotes the states at time step $t$ ($n_x$ is the number of states), i.e., the room air temperature as well as the temperatures in different layers in the walls, floor, and ceiling (all measured in $^o$C). $u^b_t \in \Rset^{n_u}$ denotes the IRA control inputs, namely heating and cooling power, ventilation, blind position, and lighting ($n_u$ is the number of actuators). The heating and cooling are represented in the thermal model as heat fluxes affecting the system states and their units are W/m$^2$, i.e., the heat fluxes are normalized by the floor area. The blind position is a number between $0$ (fully closed) and $1$ (fully open). The lighting is also normalized by the floor area and measured in W/m$^2$. $v^b_t \in \Rset^{n_v}$ denotes the disturbances that affect building's states, e.g., ambient temperature in $^o$C, solar radiation in W/m$^2$, and internal heat gains by occupants and equipment in W/m$^2$ ($n_v$ is the number of disturbances). $\Delta{u}^b_t \in \Rset^{n_r}$ denotes the uncertain change in heat fluxes due to change in power consumption of the heating/cooling devices during reserve provision, where $n_r \leq n_u$ is the number of actuators that are used for reserve provision. $R^b$ consists of the columns of $B^b_t$ that correspond to the heating/cooling actuators that provide reserves. The system's output equation is $y^b_{t} = C^bx_t^b +D^b_t u^b_t +F^bv^b_t$, where $y^b_t \in \Rset^{n_y}$ denotes the room temperature in $^o$C and illuminance in lux.

Denote by $r_{e,t}^{b,j}$ the (electric) reserve capacity of actuator $j$ of building $b$ at time step $t$. Buildings can provide \emph{up-reserves} by decreasing their consumption, and \emph{down-reserves} by increasing it\footnote{The (electric) reserve capacity is the amount of SFC reserves that becomes available to the TSO \cite{ENTSOEnetworkCode}. In the context of frequency regulation, the term up-reserves denotes increase of a generator's production or decrease of a load's consumption to increase system frequency. Similarly, the term down-reserves denotes decrease of a generator's production or increase of a load's consumption to decrease system frequency.}. For now, we assume symmetric reserve capacities; asymmetric reserves will be discussed in Section~\ref{ControllerDesign}. Since the HVAC control input for heating and cooling is defined as a heat flux, it is convenient to define also the ``thermal'' reserve capacity $r_t^{b,j}$ that has W/m$^2$ units. $r_{e,t}^{b,j}$ can be obtained from $r_t^{b,j}$  by division with the coefficient of performance (COP). For notational convenience, we use the variable $r_t^{b,j}$ in the problem formulations and call it simply reserve capacity keeping in mind that it is actually the ``thermal'' reserve capacity. In the pro-rata activation case, the reserve energy is proportional to the reserve capacity based on a normalized SFC signal \mbox{$w_t \in [-1,1]$}. Note that the reserve capacity $r_t^{b,j}$ is a decision variable for the buildings, whereas the normalized SFC signal $w_t$ is uncertain. The primitive uncertainty $w_t$ results in an uncertain change in electric power consumption of actuator $j$, $\Delta{u}_{e,t}^{b,j} = r_{e,t}^{b,j} w_t$ and the corresponding uncertain change in heat flux $\Delta{u}_t^{b,j} = r_t^{b,j} w_t$.

Denote by $x^b_{t+k|t} \in \mathbb{R}^{n_x}$ the predicted state of building $b$ for time $t+k$ at time $t$. The predicted states at time $t$ along a prediction horizon $N$ are assembled in one vector as $\x^b_t = [x^b_{t|t}~~x^b_{t+k|t}~...~x^b_{t+N|t}]^\top \in \mathbb{R}^{n_x(N+1)}$. Adopting the same notation for inputs and disturbances, the building dynamics along $N$ can be written as $\x^b_t = \A^bx^b_0 + \B_t^b \u^b_t + \E^b \v^b_t + \R^b \boldsymbol{\Delta{u}}^b_t$ and $\y^b_{t} = \C^b \x_t^b + \D^b_t \u^b_t + \F^b \v^b_t$, where the matrices $\A^b, \B^b_t$, $\E^b$, $\R^b$, $\C^b$, $\D^b_t$, and $\F^b$ are of appropriate dimensions. The constraints on outputs (thermal comfort zone) and HVAC control inputs along $N$ are

\begin{align}
\y^b_{\textrm{min}} \leq \y^b_{t} \leq \y^b_{\textrm{max}}~,~~
\u^b_{\textrm{min}} \leq \u^b_t + \Hb^b \boldsymbol{\Delta{u}}^b_t \leq \u^b_{\textrm{max}}~,  \label{yu_cons}
\end{align}
where $\Hb^b \in \mathbb{R}^{n_u \times n_r}$ has $0$ or $1$ as entries. By substituting the dynamics in \eqref{yu_cons}, the constraints can be written in terms of the control inputs and uncertainty as $\G^b \u^b_t + \Ss^b \boldsymbol{\Delta{u}}^b_t \leq \Q^b$, where the matrices $\G^b$, $\Ss^b$, and $\Q^b$ are defined as
\begin{align}
& \G^b = \begin{bmatrix}
    \G^b_p  \\
    \Ib_N \\
    -\G^b_p \\
    -\Ib_N
\end{bmatrix},
\Ss^b = \begin{bmatrix}
    \Ss^b_p \\
    \Hb^b \\
    -\Ss^b_p \\
    -\Hb^b
\end{bmatrix},
\Q^b = \begin{bmatrix}
    \y^b_{\textrm{max}}-\Q^b_p  \\ 
    \u^b_{\textrm{max}} \\
    \Q^b_p - \y^b_{\textrm{min}} \\
    -\u^b_{\textrm{min}}
\end{bmatrix}, \label{GbSbQb}\\
& \G^b_p = \C^b \B_t^b + \D^b_t~,~ \Ss^b_p = \C^b \R^b~, \label{GpbSpb} \\
& \Q^b_p = \C^b (\A^bx^b_0 + \E^b \v^b_t) + \F^b \v^b_t~, \label{Qpb}
\end{align}
and $\Ib_N$ is the N-dimensional identity matrix. Using $\G^b$, $\Ss^b$, and $\Q^b$, the HVAC input and comfort zone constraints along the prediction horizon can be represented compactly in the optimization problems of Section~\ref{ControllerDesign}.

We are interested in building aggregations for two main reasons. First, the reserve markets typically have requirements on the minimum size of the bidden reserve capacity, which is typically in the range $[1,10]$ MW, as mentioned in Section~\ref{FreqCtrl}. In most cases, individual commercial buildings cannot meet these minimum bid size requirements, so building aggregations are needed to enable participation in the reserve market. Even in markets with low minimum bid size requirements, e.g., in the range of a few hundred kWs, building aggregations would still be of interest in presence of other typical requirements, such as symmetry and minimum duration of the bidden reserve capacity. As shown in \cite{VrettosIFAC2014}, aggregating buildings with different characteristics results in a larger total reserve capacity compared with the case where each building participates individually in the market. The second argument in favor of building aggregations is more practical. An aggregator's job would be to determine the reserve capacity, bid it in the reserve market, and interact with the TSO during reserve activation and for the financial settlement. These tasks are very different to the normal activities of a building manager; therefore, the aggregator could take over this burden that would otherwise be with the building manager.

Consider an aggregation of $L$ buildings, i.e., $b=\{1 \ldots L\}$. Denote by $\x_{t}=\left[\x^1_t~~\x^2_t~...~\x^L_{t}\right]^\top \in \mathbb{R}^{n_xL(N+1)}$ the vector containing all predicted states of all buildings along $N$. Using the same notation for inputs and disturbances, the input/output constraints of the aggregation can be written as $\G \u_t + \Ss \boldsymbol{\Delta{u}}_t \leq \Q$, where $\G$, $\Ss$, and $\Q$ are block diagonal matrices with $\G^b$, $\Ss^b$, and $\Q^b$ on the diagonal, respectively. Denoting by $\w_t \in [-1,1]^{N}$ the SFC signal along $N$, and by \mbox{$\r_t \in \Rset^{n_r L N}$} a collection of the reserve capacities of all actuators and all buildings along $N$, the input/output constraints of the aggregation can be written as

\begin{align}
\tilde{\R} =
\begin{bmatrix}
    R_{1,1} & \hspace{-0.2cm} 0 & \hspace{-0.2cm} \ldots & \hspace{-0.2cm} & \hspace{-0.4cm} 0 \vspace{-0.1cm} \\
    R_{2,1} & \hspace{-0.2cm} R_{2,2} & \hspace{-0.2cm} 0 & \hspace{-0.2cm} \ldots & \hspace{-0.4cm} 0 \vspace{-0.3cm} \\
    & \hspace{-0.2cm} & \hspace{-0.2cm} \vdots \vspace{-0.3cm} \\
    R_{N,1} & \hspace{-0.2cm} R_{N,2} & \hspace{-0.2cm} \ldots & \hspace{-0.2cm} & \hspace{-0.4cm} R_{N,N}
\end{bmatrix},
\G \u_t + \Ss \tilde{\R} \w_t \leq \Q~,
\end{align}
where $\mathbf{\Delta{u}}_t = \tilde{\R} \w_t$, $\tilde{\R} \in \Rset^{\left(n_r L N\right) \times N}$ is the \emph{reserve capacity matrix}, and $R_{i,k} \in \Rset^{n_r L}$ is a column vector. The diagonal vectors of $\tilde{\R}$ are the actuators' reserve capacities for every time step $t$, i.e., $R_{t,t} = [ r_t^1 \ldots r_t^{n_rL}]^\top$. This lower triangular structure satisfies causality and allows us to model the effect of past SFC signals. For example, $R_{t,t-1}$ accounts for the signal at time $t-1$ to determine the reserve at time $t$. However, to comply with today's practice we fix $R_{i,k} = 0$ for $i \neq k$ in the following, i.e., the reserve energy depends only on the capacity and the current SFC signal. Thus, $\tilde{\R}$ is a block diagonal version of $\r_t$.

\subsection{Uncertain Frequency Signals} \label{UncertaintyModeling}
\subsubsection{Power Constraints (PC)}
In pro-rata activation \cite{swissgridAS}, the reserves are requested by the TSO via a normalized frequency signal $w_t \in [-1,1]$. This box constraint represents the power constraints of the signal. The values $w_t=-1$ and $w_t=1$ indicate full activation of up- and down-reserves, respectively. The uncertainty set along $N$ can be written as

\begin{align}
\Wset_1 = \{\w_t \in \Rset^N : ||\w_t||_\infty \leq 1 \}~. \label{Wset1}
\end{align}

\subsubsection{Power and Energy Constraints (PEC)}
To facilitate the participation of energy-constrained resources (thermal loads and storages) in AS, the frequency signals can be distinguished by their energy content. A building aggregation could then choose the preferred product to offer reserves. Energy constraints can be cast as linear constraints on the mean value of the signal as

\begin{align}
-\varepsilon T \leq \sum\nolimits_{k=t}^{t+T-1} w_k \leq \varepsilon T~, \label{EnergyCons}
\end{align}
where $\varepsilon$ is the bias coefficient, $T$ is the averaging period, and both are fixed by the TSO for a particular product. Equation \eqref{EnergyCons} implies that the bias of the signal over $T$ is bounded. If we stack \eqref{EnergyCons} along $N$, we get the polyhedral constraint on the uncertainty \mbox{$\A_w \w_t \leq \b_w$}, where $\A_w$ is a matrix with entries $-1$, $0$ or $1$, and $\b_w$ is a vector with entries $\varepsilon T$. The power constraints \eqref{Wset1} are still present, since the full reserve capacity could be requested anytime. Denote by $\I_N$ the N-dimensional identity matrix, and by $\bold{1}_N$ the N-dimensional vector with ones. Defining $\bar{\A}_w = [\A_w; \I_N; -\I_N]$ and $\bar{\b}_w = [\b_w; \bold{1}_N; \bold{1}_N]$, the uncertainty set for PEC  is

\begin{align}
\Wset_2 = \{\w_t \in \Rset^N : \bar{\A}_w \w_t \leq \bar{\b}_w \}~. \label{EnergyCons3}
\end{align}

\subsection{Other Sources of Uncertainty} \label{OtherUncert}
Apart from the SFC signal, in general, there are two other sources of uncertainty associated with our problem: (a) weather and occupancy uncertainties, i.e., deviations from the predicted values, and (b) electricity and reserve price uncertainties. For simplicity, weather/occupancy uncertainties are not considered in this paper, i.e., the predictions are assumed to be perfect. There exist several approaches in the literature of building climate control to handle these uncertainties such as stochastic MPC with additive Gaussian noise \cite{Oldewurtel2011}, scenario-based MPC \cite{Zhang2013ECC}, or simply constraint tightening, which can be relatively easily integrated in our framework. We assume that the building aggregation acquires energy from the retail market, and since the utility company tariffs are typically constant over long periods, there is no electricity price uncertainty in our case. However, the situation is different for the reserve prices. Although the reserve bid price is selected by the aggregator, there is uncertainty involved because the bid might not be accepted in the auction. Incorporating this source of uncertainty requires modeling the market clearing process and is beyond the scope of this paper, but it may well constitute an interesting topic for future research.
\begin{figure}[t]
\centering
\includegraphics[width=1\textwidth]{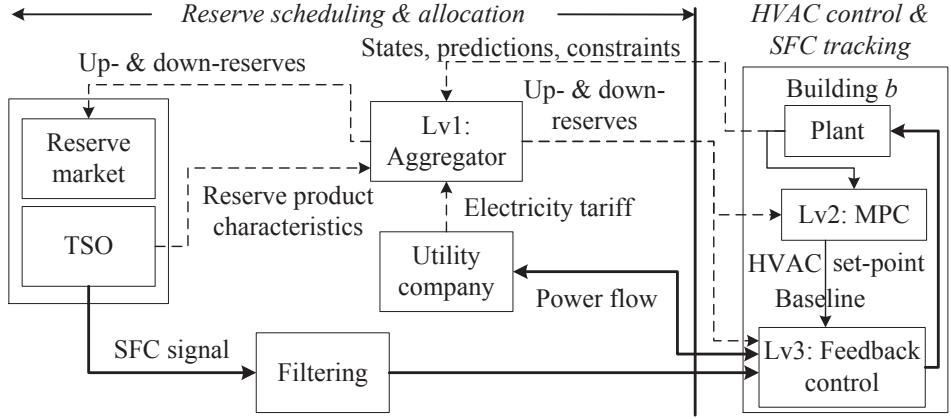}
\caption{Overview of the proposed hierarchical control scheme.} \label{fig:HierarchicalScheme}
\end{figure}

\section{Control Framework} \label{ControllerDesign}
\subsection{Hierarchical Scheme} \label{HierarchicalScheme}
In this section, we present the hierarchical control scheme for frequency reserve provision by commercial building aggregations, which is graphically shown in Fig.~\ref{fig:HierarchicalScheme}. Level $1$ (Lv$1$), the \emph{aggregator scheduling}, is performed on a daily basis centrally by an aggregator. There are two conflicting objectives in the problem: (a) minimize energy consumption through energy efficient control, and (b) deviate from the energy-optimal solution to leave slack for reserve provision. The optimal tradeoff between these conflicting objectives depends on the monetary incentives for reserves, which we call capacity payments, and the cost of electricity. Lv$1$ determines the reserve capacity, and its allocation among buildings, which achieve the optimal tradeoff while respecting occupants' comfort in case the reserve is requested. The aggregator solves a robust optimization problem using building state measurements, prices of electricity and reserves, as well as predictions of weather and occupancy. We formulate the robust problem first for PC only and second for PEC. Reliable provision of frequency reserves is critical for power system security, and so TSOs typically require $100\%$ availability of the reserve capacity \cite{swissgridAS}. For this reason, HVAC input constraints are truly hard constraints and robust optimization is a natural approach to follow.

Level $2$ (Lv$2$), the \emph{building HVAC control}, is a robust MPC \cite{MPCtextbook} that determines the energy optimal HVAC control inputs every $30$ minutes locally at each building, and leaves enough slack for reserves. Level $3$ (Lv$3$), the \emph{frequency signal filtering and tracking}, is a feedback controller that tracks the SFC signal by controlling the power consumption in real-time, e.g., every $10$ seconds. In Fig.~\ref{fig:HierarchicalScheme}, signals and control actions with thick/solid curves are real-time, the ones with thin/solid curves are executed every $30$ minutes, and the ones with dashed curves are executed once a day.

\subsection{Lv$1$: Aggregator Scheduling} \label{SchedulingLv1}
The aggregator's goal is to determine the optimal amount of reserves $\tilde{\R}^{\ast}$ to be offered in the market. Denote by $\c_t \in \Rset^{n_u L N_1}$ and $\k_t \in \Rset^{n_r L N_1}$ the electricity cost and reserve capacity payment vectors, respectively, where $N_1$ is the prediction horizon. Note that efficiency factors incorporating actuators' COP and building distribution system losses are included in $\c_t$ and $\k_t$. Lv$1$ can be cast as the robust LP
\begin{subequations} \label{GenForm}
\begin{align}
( \u_t^{\ast}, \tilde{\R}^{\ast}):= \arg\min ~ & \c_t^\top \u_t - \k^\top_t \tilde{\R} \bold{1}_{N_1} \label{ObjGenForm} \\
\text{s.t. } & \underset{\w_t \in \Wset}{\max} ( \G \u_t + \Ss \tilde{\R} \w_t )\leq \Q \label{NormalConsLv1GenForm}\\
& \M \tilde{\R} \bold{1}_{N_1} = \mathbf{0}~, \label{tender_periods_consGenForm}
\end{align}
\end{subequations}
where $\Wset$ is the uncertainty set, i.e., $\Wset \in \{\Wset_1, \Wset_2\}$. Equation \eqref{NormalConsLv1GenForm} requires that input and output constraints are satisfied even in the worst case of uncertainty realization. By appropriately selecting $\M$ in \eqref{tender_periods_consGenForm}, we can impose constant reserve capacities over a period of time and/or the block diagonal structure on $\tilde{\R}$ discussed in Section~\ref{BldModeling}. The building dynamics in \eqref{NormalConsLv1GenForm} are decoupled among buildings; however, the coupling comes via \eqref{tender_periods_consGenForm} and the objective function.

Denote by $\X_{(j)}$ the $j^\textrm{th}$ row of any matrix or vector $\X$. We derive the  robust counterpart of \eqref{GenForm} for PC and PEC. Consider the $j^\textrm{th}$ row of \eqref{NormalConsLv1GenForm} $\max_{\w_t} ( \G_{(j)} \u_t + \Ss_{(j)} \tilde{\R} \w_t )\leq \Q_{(j)}$. The term $\Ss_{(j)} \tilde{\R} \w_t$ is a scalar and can be written as $\w_t^\top \F_{(j)} \tilde{\r}$, where $\tilde{\r} \in \Rset^{n_r L N_1^2}$ is a column-wise vectorized version of $\tilde{\R}$ and $\F_{(j)} = (\I_{N_1} \otimes \Ss_{(j)}^\top)^\top \in \Rset^{N_1 \times (n_r L N_1^2)}$, where $\otimes$ is the Kronecker product. Thus, the $j^\textrm{th}$ row of \eqref{NormalConsLv1GenForm} is equivalent to

\begin{align}
\underset{\w_t \in \Wset}{\max} ( \G_{(j)} \u_t + \w_t^\top \F_{(j)} \tilde{\r})\leq \Q_{(j)}~. \label{RowNormalConsv2}
\end{align}

\subsubsection{Robust Counterpart for Power Constraints} \label{RobustCounterpartPC}
In the presence of PC only, the uncertainty set $\Wset_1$ is given by \eqref{Wset1}, i.e., $\w_t$ is constrained in an $\infty$-norm ball. In this case, we can maximize the left-hand side of \eqref{RowNormalConsv2} analytically using the dual of $\infty$-norm, i.e., the $1$-norm \cite{Lofberg2012}. Following this procedure, the deterministic equivalent of \eqref{RowNormalConsv2} is \mbox{$\G_{(j)} \u_t + ||\F_{(j)} \tilde{\r}||_1 \leq \Q_{(j)}$}. Repeating this procedure for all rows of \eqref{NormalConsLv1GenForm}, the robust counterpart problem of \eqref{GenForm} can be written as
\begin{subequations} \label{RobustCounterW1}
\begin{align}
\left ( \u_t^{\ast}, \tilde{\r}^{\ast} \right ):= \arg\min ~ & \c_t^\top \u_t - \tilde{\k}^\top_t \tilde{\r} \\
\text{s.t. } & \G_{(j)} \u_t + ||\F_{(j)} \tilde{\r}||_1 \leq \Q_{(j)} ~~ \forall j \label{RobustConsW1} \\
& \tilde{\M} \tilde{\r} = \mathbf{0}~, \label{tender_periods_RobustCounterW1}
\end{align}
\end{subequations}
where $\tilde{\k}_t \in \Rset^{n_r L N_1^2}$ and $\tilde{\M}$ are defined by $\tilde{\k}^\top_t \tilde{\r} = \k^\top_t \tilde{\R} \bold{1}_{N_1}$ and $\tilde{\M} \tilde{\r} = \M \tilde{\R} \bold{1}_{N_1}$. Problem \eqref{RobustCounterW1} is an LP, but it grows quadratically in $N_1$ since we need $N_1$ auxiliary variables and $2N_1$ additional constraints for each uncertain constraint \eqref{RowNormalConsv2} to model the $1$-norm. A similar reformulation for a general class of linear systems with reserve demands can be found in \cite{Zhang2014CDC}.

We now consider the case where either only heating or only cooling actuators provide reserves in each building, which is likely in practice to avoid energy dumping by simultaneous heating and cooling. Recall from \eqref{GbSbQb}, \eqref{GpbSpb} that $\Ss$ is a block diagonal collection of $\Ss^b = [\C^b \R^b; \Hb^b; -\C^b \R^b; -\Hb^b]$. If only heating (cooling) actuators are used for reserve provision, then all entries of $\R^b$ are non-negative (non-positive). Additionally, all entries of $\C^b$ and $\Hb^b$ are non-negative by construction. Therefore, every row of $\Ss$ contains either only non-negative or only non-positive entries and, by construction, the same holds for all entries of $\F_{(j)}$, i.e., $[ \F_{(j)}(i,k) \geq 0 ~\forall i,k ]$ or $[ \F_{(j)}(i,k) \leq 0 ~\forall i,k ]$ $\forall j$. Based on the definitions of $\F_{(j)}$, $\Ss_{(j)}$, $\tilde{\r}$, and $\r_t$, and recalling that $\tilde{\r}$ is non-negative, $||\F_{(j)} \tilde{\r}||_1$ can be equivalently written as the linear term

\begin{align}
&||\F_{(j)} \tilde{\r}||_1 = \sum\nolimits_{i=1}^{N_1} \sum\nolimits_{k=1}^{n_{\textrm{el}}} | \F_{(j)}(i,k) \tilde{\r}(k) | = \label{Derivation}  \\
& \sum\nolimits_{i=1}^{N_1} \sum\nolimits_{k=1}^{n_{\textrm{el}}} |\F_{(j)}(i,k)| \tilde{\r}(k)
 = \bold{1}_{N_1} |\F_{(j)}|_{\star} \tilde{\r} = |\Ss_{(j)}|_{\star} \r_t, \nonumber
\end{align}
where $n_{\textrm{el}}=n_r L N_1^2$ and $|\cdot|_{\star}$ denotes the element-wise absolute value operator. In this case, the more general formulation \eqref{RobustCounterW1} can be simplified to the following LP that has the same size as \eqref{GenForm}, and so can be solved efficiently\footnote{Note that \eqref{RobustCounterW1simple} is a special case of \eqref{RobustCounterW1}, and it is equivalent to formulation (13)-(15) in our previous work \cite{VrettosIFAC2014}.}.

\begin{subequations} \label{RobustCounterW1simple}
\begin{align}
\left ( \u_t^{\ast}, \r_t^{\ast} \right ):= \arg\min ~ &\c_t^\top \u_t - \k^\top_t \r_t \\
\text{s.t. } & \G_{(j)} \u_t + |\Ss_{(j)}|_{\star} \r_t \leq \Q_{(j)} ~~ \forall j \\
& \M \r_t = \mathbf{0}~. \label{tender_periods_RobustCounterW1simple}
\end{align}
\end{subequations}

\subsubsection{Robust Counterpart for Power and Energy Constraints}
With PEC, the uncertainty set $\Wset_2$ is a polyhedron and duality theory can be applied to derive the robust counterpart problem \cite{BoydConvex,Lofberg2012}. We write constraint \eqref{RowNormalConsv2} as an optimization problem over $\w_t$

\begin{subequations} \label{RobustConsW2part1}
\begin{align}
& \underset{\w_t}{\max} & ( \G_{(j)} \u_t + \w_t^\top \F_{(j)} \tilde{\r} )\leq \Q_{(j)} \\
& \text{s.t. } & \bar{\A}_w \w_t \leq \bar{\b}_w~.
\end{align}
\end{subequations}
By deriving the dual problem of \eqref{RobustConsW2part1} for all $j$, we obtain the following robust counterpart of \eqref{GenForm}

\begin{subequations} \label{RobustCounterW2}
\begin{align}
( \u_t^{\ast}, \tilde{\r}^{\ast}, \boldsymbol{\lambda}_{(j)}^{\ast} ):= &\arg\min ~\c_t^\top \u_t - \tilde{\k}^\top_t \tilde{\r} \\
\text{s.t. } & \bar{\b}_w^\top \boldsymbol{\lambda}_{(j)} + \G_{(j)} \u_t - \Q_{(j)} \leq 0~~~\forall j \\
& \bar{\A}_w^\top \boldsymbol{\lambda}_{(j)} = \F_{(j)} \tilde{\r}~~~\forall j  \\
& \boldsymbol{\lambda}_{(j)} \geq \mathbf{0}~\forall j,\\
&\tilde{\M} \tilde{\r} = \mathbf{0}~, \label{tender_periods_RobustCounterW2}
\end{align}
\end{subequations}
where $\boldsymbol{\lambda}_{(j)}$ is the vector of dual variables. Problem \eqref{RobustCounterW2} is an LP, but its size grows quadratically in $N_1$. Although this dualization technique holds also for PC, we apply analytic maximization in that case since it results in fewer variables and constraints.

\emph{Remark:} Asymmetric reserves can be modeled defining $\tilde{\R}^{+}$, $\w_t^{+} \in [0,1]^{N_1}$ for down- and $\tilde{\R}^{-}$, $\w_t^{-} \in [-1,0]^{N_1}$ for up-reserves. In case of PC only, the uncertainty set remains a polyhedron and a tractable robust counterpart problem can be derived using analytic maximization. The reader is referred to \cite{VrettosIFAC2014} for detailed formulations. In case of PEC, the non-linear constraint $w_t^{-} w_t^{+}=0$ is needed to ensure that up- and down- reserves are not requested simultaneously. Therefore, the uncertainty set is not a polyhedron any more and the dualization technique cannot be applied in this case to derive the robust counterpart problem.

\subsection{Lv$2$: Building HVAC Control} \label{BuildingControl}
Given the optimal reserve allocation from Lv$1$, in Lv$2$ the HVAC control inputs are determined every $30$ minutes by the robust MPC with prediction horizon $N_2$

\begin{subequations} \label{GenFormLv2}
\begin{align}
{\u}_t^{b,\ast}:= &\arg\min~~ {(\c^b_t)}^\top\u^b_t \\
\text{s.t. } &\underset{\w_t \in \Wset}{\max} \left( \G^b \u^b_t + \Ss^b \tilde{\R}^{b,\ast} \w_t \right) \leq \Q^b \label{GenFormLv2NormalCons}~,
\end{align}
\end{subequations}
where $\c^b_t$ and $\tilde{\R}^{b,\ast}$ are the parts of $\c_t$ and $\tilde{\R}^{\ast}$, respectively, that are relevant for building $b$. The first input of the optimal control sequence of \eqref{GenFormLv2} determines the Lv$2$ setpoint of the HVAC system for the next $30$ minutes\footnote{Although a consumption schedule is calculated in Lv$1$, the MPC of Lv$2$ can reduce the costs due to less uncertainty (recent reserve requests are known and better weather forecasts might be available) and, possibly, shorter optimization time steps. In case of plant-model mismatches, MPC additionally reduces constraint violations due to its closed-loop operation. The MPC schedule is the building's baseline consumption, and is communicated to the aggregator. Since the baseline is a by-product of the formulation, baseline prediction methods that have proven to be hard \cite{kiliccote_field_2012} are not required. Furthermore, the predictive nature of MPC inherently accounts for rebound effects due to reserve provision when calculating future HVAC setpoints \cite{pietteLBNL2012}.}, $u_t^{b,\text{Lv$2$}}$. Problem \eqref{GenFormLv2} formulates an MPC with open-loop predictions, i.e., the optimization is performed explicitly over the control inputs $\u^b_t$. MPC with closed-loop predictions, i.e., optimization over affine policies of the uncertainty, showed minor or zero performance improvement in this case, and so it is not used. In the following, we derive the robust counterparts of \eqref{GenFormLv2} for PC and PEC.

\subsubsection{Robust Counterpart for Power Constraints}
In this case, the deterministic equivalent of \eqref{GenFormLv2} can be obtained by substituting \eqref{GenFormLv2NormalCons} with $\G_{(j)}^b \u^b_t \leq \Q_{(j)}^b - ||\F_{(j)}^b \tilde{\r}^{b,\ast}||_1~\forall j$, where $\F_{(j)}^b$ is defined similarly to $\F_{(j)}$ but for a single building, and $\tilde{\r}^{b,\ast}$ is a column-wise vectorized version of $\tilde{\R}^{b,\ast}$. Note that $\tilde{\r}^{b,\ast}$ is fixed in Lv$2$, and the right-hand side of the inequality is a constant.

\subsubsection{Robust Counterpart for Power and Energy Constraints}
With PEC, the robust counterpart of \eqref{GenFormLv2} is
\begin{subequations} \label{RobustCounterLv2W2}
\begin{align}
( \u_t^{b,\ast}, \boldsymbol{\lambda}_{(j)}^{b,\ast} ):= &\arg\min {(\c^b_t)}^\top \u^b_t \\
\text{s.t. } & \bar{\b}_{w,t}^\top \boldsymbol{\lambda}_{(j)}^b + \G_{(j)}^b \u^b_t - \Q_{(j)}^b \leq 0~~~\forall j \\
& \bar{\A}_{w,t}^\top \boldsymbol{\lambda}_{(j)}^b = \F_{(j)}^b \tilde{\r}^{b,\ast}~\forall j,\\
&\boldsymbol{\lambda}_{(j)}^b \geq \mathbf{0}~\forall j~. \label{EqConsLv2W2}
\end{align}
\end{subequations}
Problem \eqref{RobustCounterLv2W2} is similar to \eqref{RobustCounterW2}, but there are two main differences: first, $\tilde{\r}^{b,\ast}$ is fixed; and second, $\bar{\A}_{w,t}$ and $\bar{\b}_{w,t}$ are time-varying. For a time step $t$ in the averaging interval $[t_1,t_2]$ of length $T$, \eqref{EnergyCons} can be written as $-\varepsilon T - w_{\textrm{p},t} \leq \sum_{k=t}^{t_2} w_k \leq \varepsilon T - w_{\textrm{p},t}$, where $w_{\textrm{p},t} = \sum_{k=t_1}^{t-1} w_k$ is known because the uncertainty up to $t-1$ is realized. Thus, the coupling constraint on $\{w_t, \ldots, w_{t_2}\}$ depends on the energy content of the SFC signal in the previous time steps of the averaging interval.

\subsection{Lv$3$: Frequency Signal Filtering and Tracking} \label{SFCtracking}
In Lv$3$, the HVAC consumption is controlled around $u_t^{b,\text{Lv$2$}}$ for reserve provision. We consider water-based HVAC systems that are common in Europe, but the proposed reserve scheduling problem applies also to air-based systems \cite{MaasoumyACC2014paper1}. The power consumption of water circulation pumps in water-based HVAC systems is typically small, and so we directly control the heating or cooling devices, e.g., heat pumps (HPs), to provide reserves. The desired power consumption of HP $j$ of building $b$ is

\begin{align} \label{Lv3}
u_t^{b,j,\text{Lv$3$}} &= u_t^{b,j,\text{Lv$2$}} + \Delta{u}^{b,j}_t = u_t^{b,j,\text{Lv$2$}} + w_t r^{b,j,\ast}_t~,
\end{align}
where $r^{b,j,\ast}_t$ is fixed from Lv$1$. In case of asymmetric reserves, $r^{b,j,\ast}_t$ is equal to the down-reserve capacity if $w_t \geq 0$, and equal to the up-reserve capacity if $w_t < 0$. In a fast time scale, and depending on the HVAC system, the HP consumption can be controlled by modifying either the water temperature setpoint at condenser's outlet or the refrigerant's flow rate via valves. We rely on the second approach that was experimentally shown to be able to track fast reference signals, e.g., SFC signals, in \cite{Hagiwara2011}. The desired HP consumption $u_t^{b,j,\text{Lv$3$}}$ can be tracked using a feedback PI controller.

With PC only, the normalized reserve request $w_t$ sent from the TSO is the original SFC signal, whereas with PEC, $w_t$ is a filtered version of that signal. With reasonable modifications, the current operational paradigm at the TSO side can integrate multiple SFC signals with different energy contents. In the current paradigm, there is a single SFC signal that is the output of a PI controller with the area control error (ACE) as input. In the new paradigm, the TSO provides a number of reserve products and each reserve provider chooses the product to offer its reserve capacity. Assume that the total reserve request (original SFC signal) at time step $t$ is $\tilde{w}_t$. The TSO will decompose $\tilde{w}_t$ into the desired number of signals (e.g., using a filter bank) and will sent the appropriate components to the providers depending on the reserve product they offer reserves for. To simulate this process in this paper, we use a causal Chebyshev filter to get the energy-constrained signal $w_t$ sent to the building aggregation, but other filters might also be used. Similar filtering approaches have been used in previous works on power system applications, e.g., \cite{AvramiotisGM2014}. Note that although $w_t$ is a filtered signal, it does not mean that its bias is zero but instead that its bias is bounded. The filter's transfer function is

\begin{align} \label{Chebyshev}
H(z) = \frac{\sum_{i=0}^{n_{\textrm{f}}} b_i z^{-i}}{1+\sum_{i=1}^{n_{\textrm{f}}} a_i z^{-i}}~,
\end{align}
where $n_{\textrm{f}}$ is the filter's order, and $a_i$, $b_i$ are its coefficients that depend on the pass-band edge frequency $f_\textrm{c}$. From a TSO perspective, $n_{\textrm{f}}$ and $f_\textrm{c}$ can be chosen such that the resulting low-frequency component (LF signal) and high-frequency component (HF signal) have the desired ramping rates and energy contents. In this paper, we fix $n_{\textrm{f}}=3$ since it showed good performance in preliminary simulations.

It is important to clarify how the SFC signal filtering is taken into account in the higher levels of the control hierarchy. Recall that the filtered signal is tracked every few seconds in Lv$3$, whereas the decisions in Lv$1$ (reserve scheduling) and Lv$2$ (determination of optimal building setpoints) are taken every $30$ minutes. Due to this time scale separation, the important information from the filtering of Lv$3$ that needs to be conveyed to Lv$1$ and Lv$2$ is only the integral of the SFC signal over this period, i.e., the bias of the signal. This is needed for example in Lv$1$ to schedule the reserve capacities without violating the buildings' thermal comfort constraints. By formulating the energy constraint of the SFC signal as the linear inequalities \eqref{EnergyCons}, we can directly account for the signal's bias in the optimization problems of Lv$1$ and Lv$2$ in a tractable way. Note that the bias coefficients $\varepsilon$ for different averaging periods $T$ in \eqref{EnergyCons} can be empirically obtained from the filter \eqref{Chebyshev}. To do so, one can simply apply the filter on historical data of SFC signals using different $f_\textrm{c}=1/T$.

Due to the robust design, any admissible reserve request $w_t$ that satisfies \eqref{EnergyCons} will not lead to comfort zone or input constraint violations, provided that there is no significant plant-model mismatch. The tracking quality of $w_t$ depends on HP's mechanical delays and dead-times, ramping limits, and minimum down-times and/or run-times. If such dynamics are significant, tests similar to the ones in \cite{Hagiwara2011} must be performed to identify upper limits on the frequency content of $w_t$ that result in good tracking by the HP. In this case, the building aggregation could form a coalition with faster resources, e.g., an aggregation of residential electric water heaters, exclude very high frequency components of the SFC signal using an appropriate band-pass filter, and send them to the faster resources. Since the focus of this paper is on the robust reserve scheduling and allocation side, we assume that such HP dynamics are negligible, i.e., the reference $u_t^{b,j,\text{Lv$3$}}$ (and so the SFC signal) can be perfectly tracked in our simulations if the comfort zone and input constraints are satisfied.

\section{Performance of Control Framework} \label{ControllerPerformance}
\begin{table}[t]
\renewcommand{\arraystretch}{1.1}
\caption{Bias Coefficients ($\varepsilon$) for the SFC Signal and its High-frequency Component (HF) for Different Averaging Periods ($T$)}
\centering
\begin{tabular}{ccccccc}
\hline
$T$ & $1$ h & $2$ h & $4$ h & $6$ h & $8$ h & $12$ h\\
\hline
SFC $2009$ & $1.000$ & $0.989$ & $0.952$ & $0.796$ & $0.592$ & $0.505$\\
HF $2009$ & $0.528$ & $0.467$ & $0.337$ & $0.273$ & $0.384$ & $0.229$\\
\hline
SFC $2012$ & $0.927$ & $0.781$ & $0.674$ & $0.624$ & $0.553$ & $0.448$\\
HF $2012$ & $0.382$ & $0.300$ & $0.317$ & $0.290$ & $0.237$ & $0.203$\\
\hline
\end{tabular}
\label{tab:LFCanalysis}
\end{table}

\subsection{Investigation Setup} \label{InvestigationSetup}
We investigate the performance of the proposed control framework in simulations with an aggregation of typical Swiss office buildings. We consider two HVAC systems: in system A, heating is performed via radiators with coefficient of performance (COP) equal to $3$, whereas cooling with cooled ceilings (COP $= 3.5$); in system B, both heating and cooling are performed using thermally activated building systems (TABS) with COP $= 3.4$. We also differentiate between heavy (eh) or light (el) building envelope, high (wh) or low (wl) window area fraction, and high (gh) or low (gl) internal gains. In our simulations, we consider an aggregation of $6$ large buildings ($15000$ m$^2$ each) with the following characteristics: $\textrm{A}1$ = \{A,eh,wh,gh\}, $\textrm{A}2$ = \{A,eh,wl,gl\}, $\textrm{A}3$ = \{A,el,wl,gl\}, $\textrm{B}1$ = \{B,eh,wh,gh\}, $\textrm{B}2$ = \{B,eh,wl,gl\}, and $\textrm{B}3$ = \{B,el,wl,gl\}. Typical occupancy profiles were used, whereas weather data were provided by Meteoswiss (the Swiss federal office of meteorology and climatology). More information regarding the buildings can be found in \cite{Oldewurtel2011,VrettosIFAC2014}. Because of the thermal inertia, heating/cooling actuators can be used to provide frequency reserves. In buildings with the HVAC systems considered here, the fresh air flow rate is usually kept constant because changes would be immediately realized by the occupants. For this reason, we do not use ventilation for reserve provision. The temperature comfort zone during working hours is $[21,24]^o$C in winter and $[22,25]^o$C in summer, which is in accordance with the ASHRAE 55-2013 standard \cite{CBEcomforttool}. During non-working hours and weekends, the comfort zone is relaxed to $[12,35]^o$C in both seasons. The optimizations are performed with a time step of $30$~minutes, which is the discretization step of building models \eqref{OriginalModel1}, and the prediction horizons of Lv$1$ and Lv$2$ are fixed to $N_1 = 96$ ($2$~days) and $N_2 = 48$ ($1$~day), respectively. We assume symmetric, daily reserves, i.e., constant reserve capacity over a day, and capacity payments $10\%$ higher than the electricity price, i.e., $k = 1.1 c$.

\subsection{Parameters of Energy Constraints} \label{EnergyConsParams}
To apply energy constraints as in \eqref{EnergyCons}, we determine appropriate pairs of averaging period and bias coefficient $(T, \varepsilon)$ based on the historical normalized SFC signals $w_k \in [-1,1]$ from the Swiss control area for $2009$ and $2012$. We consider six different averaging periods, $T=\{1,2,4,6,8,12\}~\textrm{hours}$, and calculate six sets of filter parameters in \eqref{Chebyshev}, one for each of the pass-band edge frequencies $f_\textrm{c}=1/T$. For each value of $T$, we filter the historical SFC signals using the corresponding filter \eqref{Chebyshev} to get the HF signals for $2009$ and $2012$. For each of the four signals and for each value of $T$, we calculate $\varepsilon$ as the largest absolute mean value of the SFC or HF signal over any period $T$, i.e., $\varepsilon = \max(|(1/T) \cdot \sum_{k=1}^{T} w_k|)$. The results are summarized in Table~\ref{tab:LFCanalysis}. Notice that the original SFC signals can be significantly biased over periods of several hours. Note also that the biases of the HF signals are significantly lower than those of the original SFC signals, and that the signals in $2012$ are generally less biased than in $2009$. This is because from March $2012$ the ACE of Switzerland is netted with that of other European countries before generating the SFC signal. We use the HF signal of $2012$ and select $(T=2~\textrm{h}, \varepsilon=0.3)$ for the simulations of this section, according to Table~\ref{tab:LFCanalysis}.

\subsection{Operation for Typical Weeks} \label{TypicalOperation}
\begin{figure}[t]%
\centering
\begin{subfigure}
\centering
\includegraphics[width=1\textwidth,height=2in]{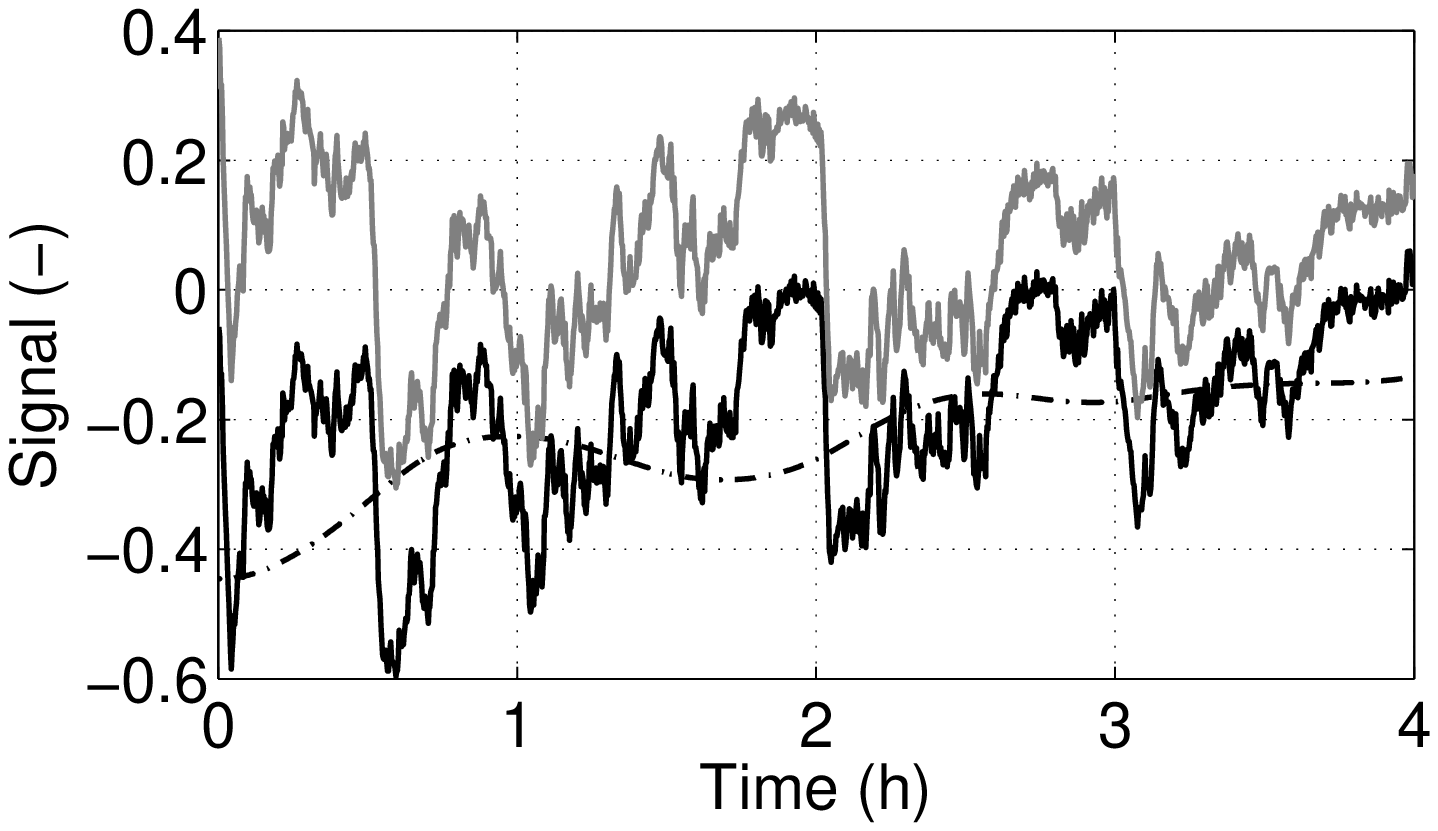}
\caption{A $4$-hour extract of the original SFC signal, its low-frequency (LF), and its high-frequency (HF) components for a sample day.}
\label{fig:LFCsignals1}
\end{subfigure}
\begin{subfigure}
\centering
\includegraphics[width=1\textwidth,height=2in]{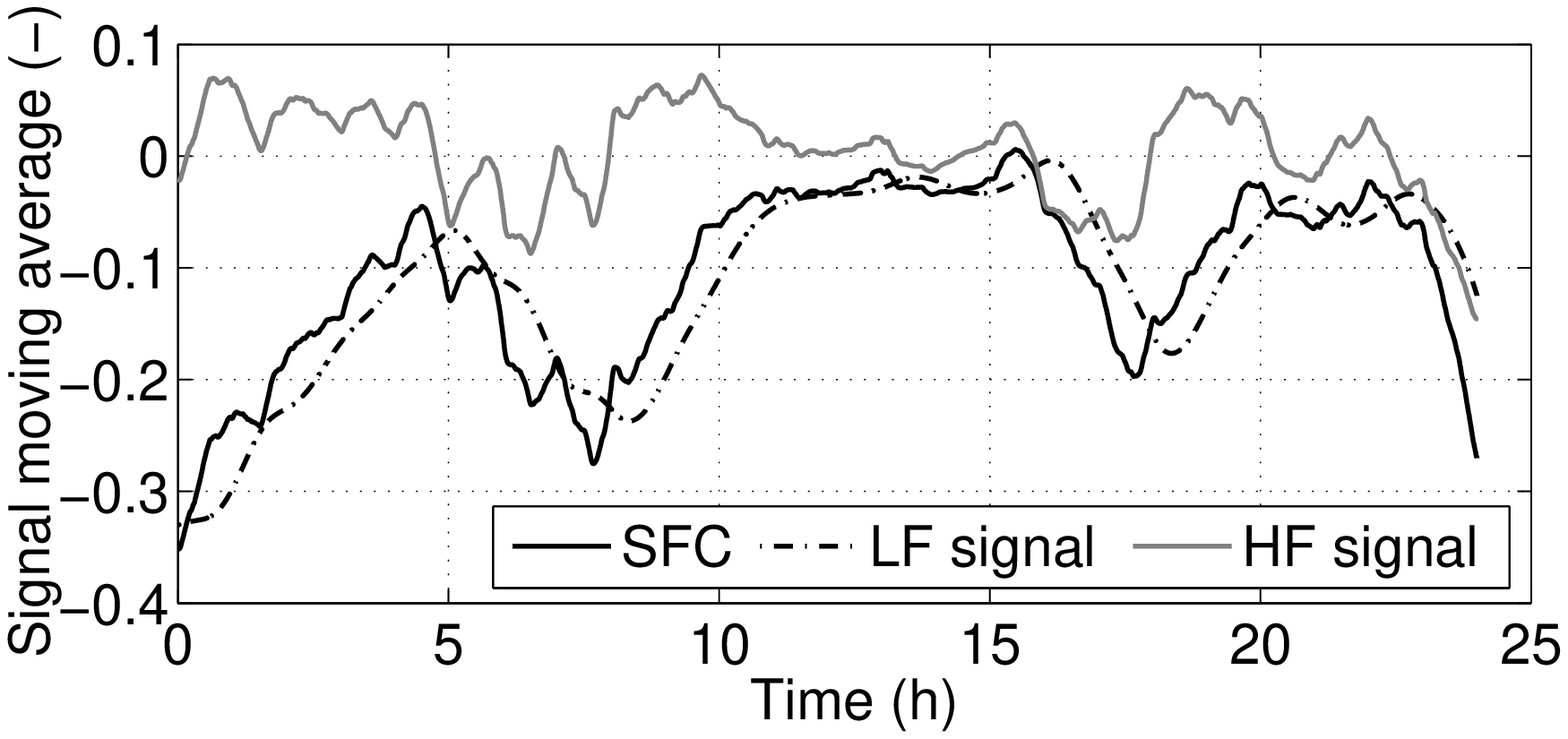}
\caption{The moving averages of the signals of Fig.~\ref{fig:LFCsignals1} for the whole sample day.}
\label{fig:LFCsignals2}
\end{subfigure}
\end{figure}

We present simulation results for typical weeks in winter and summer. In Fig.~\ref{fig:LFCsignals1}, we show a \mbox{$4$-hour} extract of the SFC, LF, and HF signals for a sample day, whereas in Fig.~\ref{fig:LFCsignals2} we show the $2$-hour moving averages of the signals for the whole day. Although the original SFC signal is mostly negative, its bias is absorbed by the LF signal, and so the HF signal is approximately zero-mean. The bias $\varepsilon$ of the SFC signal is larger than $0.3$, whereas the energy-constrained HF signal has a bias less than $0.15$.

\begin{figure}[t]
\centering
\begin{subfigure}
\centering
\includegraphics[width=1\textwidth]{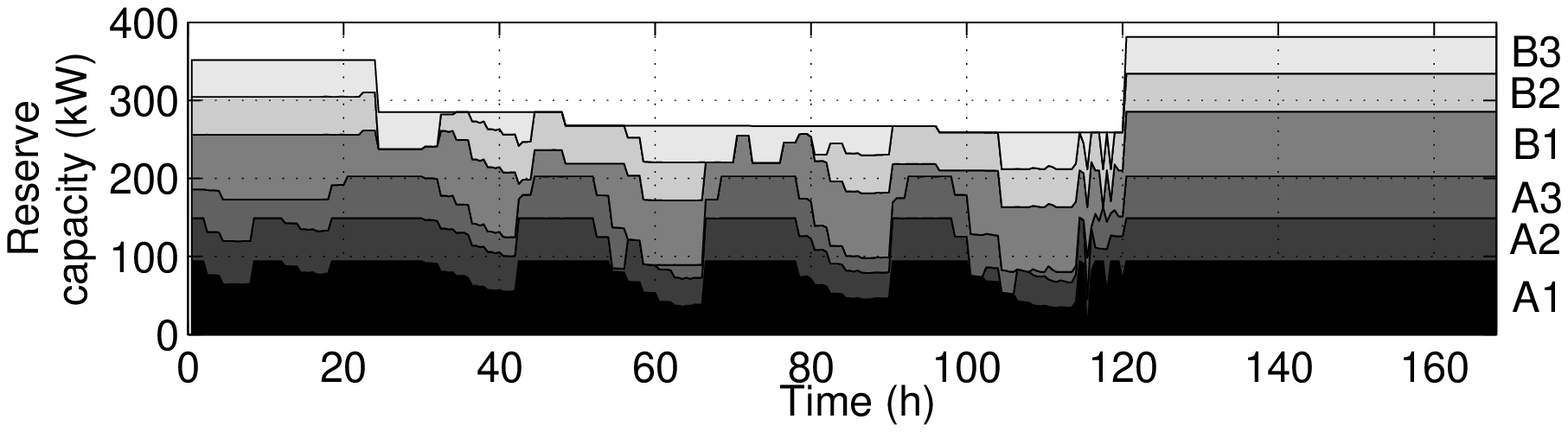}
\caption{Optimal reserve allocation among buildings (winter week).} \label{fig:ReserveCapacitiesWinter}
\end{subfigure}
\begin{subfigure}
\centering
\includegraphics[width=1\textwidth]{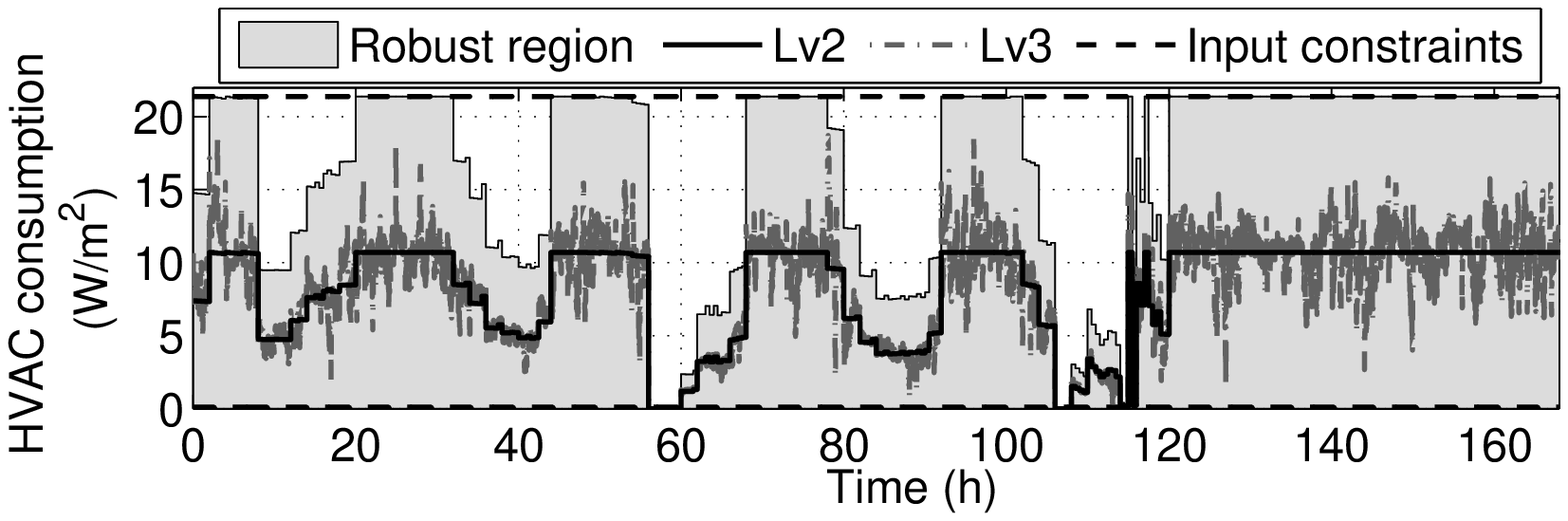}
\caption{HVAC heating power for building A$3$ (winter week).} \label{fig:InputsWinter}
\end{subfigure}
\begin{subfigure}
\centering
\includegraphics[width=1\textwidth]{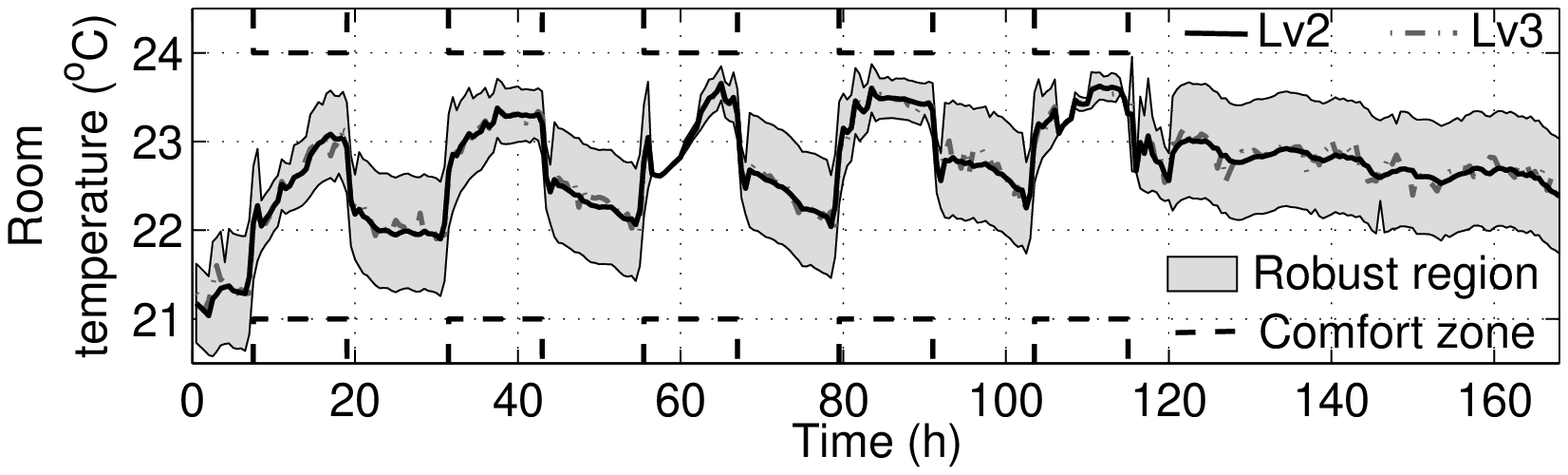}
\caption{Room temperature for building A$3$ (winter week).} \label{fig:RoomTemperatureWinter}
\end{subfigure}
\end{figure}

Figure~\ref{fig:ReserveCapacitiesWinter} shows the optimal total reserve capacity of the aggregation and its allocation among buildings for the winter week. The capacity is constant for each day and ranges from approximately $260$~kW on Friday to $380$~kW on Saturday with an average weekly value of $313$~kW. Note that the capacity is shifted among buildings in a way that maximizes the total capacity of the aggregation. Interestingly, the buildings offer higher reserve capacities when they normally consume less power. For example, buildings with system A contribute mainly at night, whereas buildings with system B offer more reserves during working hours, because they prefer to preheat at night. The scheduled total energy consumption of the aggregation is spread throughout the whole week to maximize the reserve potential. During the weekend, the buildings are not occupied and the comfort zone is larger, which results in higher reserve capacities compared to working days.

The optimal operation of building A$3$ based on Lv$2$ and Lv$3$ is presented in Figs.~\ref{fig:InputsWinter} and~\ref{fig:RoomTemperatureWinter}. Figure~\ref{fig:InputsWinter} shows the HVAC system heating power and Fig.~\ref{fig:RoomTemperatureWinter} the resulting temperature trajectories. The black-solid curves indicate the scheduled consumption and temperature by the MPC, whereas the grey-dashed curves correspond to the final values after tracking the HF signal. The grey envelope shows the problem's robust region, which is defined as the region of the thermal comfort zone with the following property. If the temperature is within the robust region at time step $t=0$, then for any SFC signal with the appropriate energy content there exists a feasible control input trajectory that satisfies the thermal comfort and HVAC input constraints (the black-dashed curves in both figures) along the prediction horizon $t \in [0,N-1]$. In our simulations, the HVAC inputs and room temperature stay always within the robust region.

In summer, the reserve potential from cooling actuators is also significant. During the considered week, the capacities range from approximately $60$~kW on Sunday to $420$~kW on Saturday with an average weekly value of $323$~kW. We observed a more uniform distribution of reserves among different buildings in summer, compared to winter, because cooled ceilings and TABS have similar time constants. Due to space limitations, plots similar to Figs.~\ref{fig:ReserveCapacitiesWinter},~\ref{fig:InputsWinter} and~\ref{fig:RoomTemperatureWinter} are omitted.

Based on our results, the considered $6$ buildings with floor area $15000$ m$^2$, average rated heating power $27$ W/m$^2$, and average rated cooling power $32$ W/m$^2$ provide a reserve capacity equal to $313$ kW and $323$ kW on average in winter and summer, respectively. Using simple linear extrapolation, we find that the required minimum reserve capacity of $5$ MW in Switzerland can be provided by $96$ (similar) buildings in winter and $93$ buildings in summer, i.e., approximately $100$ buildings in both seasons. Note that this is only a rough estimate as: (a) it is based on the average reserve capacity values; and (b) the extrapolation might overestimate the required number of buildings to meet the $5$ MW limit, because the larger the aggregation the more flexibility exists in allocating reserves among buildings.

\subsection{Effect on Energy Consumption} \label{EnergyConsumption}
To maximize the reserve potential, the buildings operate in a less energy efficient way. For example, during the considered winter week reserve provision resulted in a consumption $60\%$ higher than that of energy efficient building control, i.e., an MPC with objective to minimize electricity cost without offering reserves. As shown in Fig.~\ref{fig:InputsWinter}, in order to provide reserves in both directions, the buildings try to operate close to the middle of the heating/cooling device's power range. This is in contrast to an energy efficient operation, where the power consumption would be as close as possible to the minimum value. However, the increase in energy consumption does not mean that the building control is suboptimal. For the given electricity price and capacity payment, this building operation minimizes the total cost defined as the sum of electricity cost and reserve profit.

It is important to note that the $60\%$ consumption increase is in comparison with an MPC-based energy efficient building control. However, most buildings today operate with supervisory rule-based controllers, i.e., not optimal controllers. Therefore, if the proposed methods are applied to such buildings, the observed increase in consumption is expected to be less than the reported value. Although energy efficiency is usually the goal in building control, increasing the energy consumption is not necessarily a drawback if this helps accommodate more RES in the power system, and at the same time the additional electric energy is stored as thermal energy in the buildings.

\subsection{Discussion on Imperfect Disturbance Predictions} \label{DiscussImperfectPred}
As mentioned in Section~\ref{OtherUncert}, the formulations and simulations in this paper assumed perfect weather and occupancy predictions. The robust formulation guarantees satisfaction of the HVAC and comfort constraints for any admissible reserve request and perfect weather and occupancy predictions, but it cannot provide mathematical guarantees on the worst case temperature deviations in case of imperfect predictions. Empirical guarantees could be provided by Monte Carlo simulations: (i) schedule the reserve capacities using imperfect disturbance predictions; (ii) simulate the building operation under disturbance uncertainty, i.e., the weather and occupancy realizations are different to the predictions; and (iii) analyze the results to keep track of the number and magnitude of temperature deviations. Alternatively, probabilistic guarantees could be obtained by modeling the weather and occupancy uncertainty via scenarios and then robustifying against the reserve uncertainty separately for each scenario. This is an exciting research direction for future work.

However, some intuition can be provided without following any of the above two approaches. Consider a building in energy efficient operation using a deterministic MPC that relies on an imperfect weather forecast. As explained in Section~\ref{EnergyConsumption}, the building would operate close to the minimum power consumption, and thus the temperature trajectory would stay close to one of the comfort zone boundaries. Now consider the same building operated under reserve provision. As shown in Fig.~\ref{fig:RoomTemperatureWinter}, the building would operate closer to the middle of the comfort zone to maximize the reserve capacity. For this reason, we expect the reserve provision case to result in smaller and perhaps less frequent comfort zone violations compared to the energy efficient control case, if the building is exposed to the same weather and occupancy uncertainty in both cases.

\section{Sensitivity Analysis} \label{SensitivityAnalysis}
\subsection{Capacity Payments} \label{CapPayments}
\begin{figure}[t]%
\centering
\begin{minipage}{0.5\linewidth}%
\centering \includegraphics[width=2.5in]{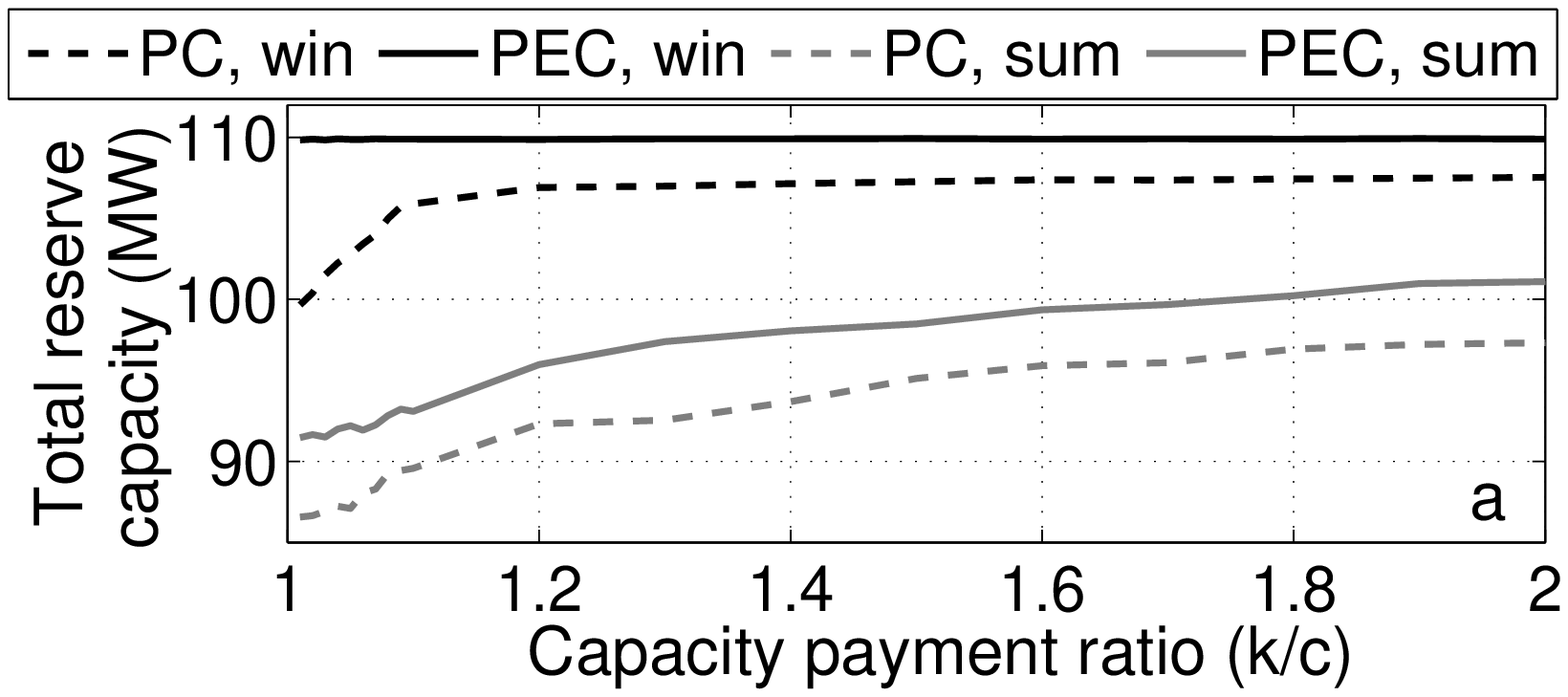}
\end{minipage}%
\begin{minipage}{0.5\linewidth}%
\centering \includegraphics[width=2.5in]{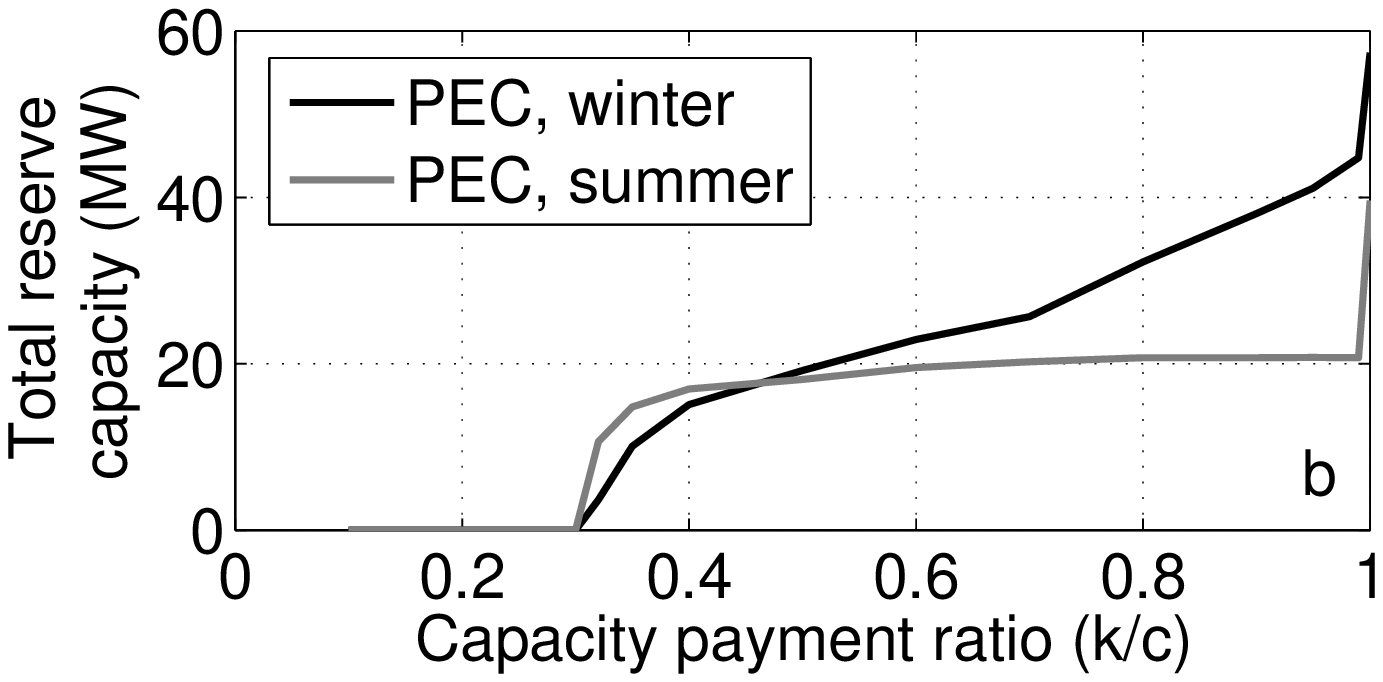}
\end{minipage}%
\caption{Bid curves of building aggregation in winter and summer. Left: for PC and PEC, and ratios $k/c>1$. Right: for PEC and ratios $k/c \leq 1$.}%
\label{fig:CapacityPayment}%
\end{figure}

As explained in Section~\ref{HierarchicalScheme}, the proposed reserve scheduling methods identify the optimal tradeoff between minimizing energy consumption and leaving enough slack for reserve provision. The buildings would not deviate from the energy efficient control and would not offer any reserves if the additional electricity cost occurring due to this deviation were higher than the reward received for the slack provided as reserve capacity. In principle, the amount of reserves depends on the relationship (ratio) between the capacity payment $k$, i.e., the remuneration for each kW of reserve capacity provided, and the electricity price $c$, i.e., the cost for each kWh of electricity consumed. In this Section, we investigate this relationship by running simulations, similar to the ones in Section~\ref{TypicalOperation}, over $2$-week periods in winter and summer and for various $k/c$ ratios. The total reserve capacity for each case, i.e., the sum of the capacities of each day of the $2$-week period, is presented in Fig.~\ref{fig:CapacityPayment}, where the left plot is for ratios $k/c > 1$ and the right plot is for ratios $k/c \leq 1$. These plots represent the aggregation's bid curves because they communicate how much capacity the aggregation is willing to bid in the reserve market depending on the payment it receives for each kW of the capacity.

For $k/c >1$ (left plot - Fig.~\ref{fig:CapacityPayment}a) the simulations are performed for winter and summer with and without consideration of energy-constrained SFC signals: the black curves correspond to winter weeks (``win'') and the grey curves to summer weeks (``sum''), whereas the dashed curves are for PC and the solid curves for PEC. Our simulations show that with the same financial incentive and for both seasons, the buildings are willing to offer up to $10\%$ more reserves compared with PC if energy constraints are considered. Note that the gap between PC and PEC is generally larger for lower $k/c$, particularly in winter. In winter, the capacity saturates at its maximum value at $k/c=1.1$, whereas in summer it increases monotonically as the ratio increases up to $2$.

The analysis of Fig.~\ref{fig:CapacityPayment}a focused on $k/c>1$, which is a necessary condition for reserve provision with PC. This observation was also made in our previous work \cite{VrettosIFAC2014}; however, here we provide an explanation by studying the structure of problem \eqref{RobustCounterW1}. If $k<c$, the optimal solution is $(\u_t^{\ast}, \tilde{\r}^{\ast})=(\u_t^{\textrm{min}},\bold{0})$, where $\u_t^{\textrm{min}}$ is the energy optimal scheduling. If $k=c$, the optimal cost is $0$ and any solution within the feasible range of $\tilde{\r}$ will be optimal. If $k>c$, the optimal solution is $(\tilde{\r}^{\textrm{max}},\tilde{\r}^{\textrm{max}})$, where $\tilde{\r}^{\textrm{max}}$ is the upper limit of $\tilde{\r}$, and the optimal cost will be $(\c_t^\top-\k^\top_t) \tilde{\r}^{\textrm{max}}<0$, i.e., the aggregation earns profit. The limit $\tilde{\r}^{\textrm{max}}$ depends on input/output constraints, and so different solutions are obtained for different $k/c>1$ ratios, as shown in Fig.~\ref{fig:CapacityPayment}a. In case of daily reserves, $k/c>1$ need not to be satisfied point-wise throughout the whole day; instead, $k/c<1$ can be chosen during daytime, and $k/c>1$ at night when electricity prices might be lower. Reserve provision will be triggered if $||k||_1/||c||_1>1$, i.e., the capacity payment is on average higher than the electricity price.

On the other hand, with energy-constrained SFC signals (PEC) reserves can be provided also with ratios $k/c<1$ throughout the whole day. It is easier to explain this with an example. Assume that the buildings have declared a capacity $r(t)$ for day $d$. Assume also that up regulation (i.e., consumption decrease) is mainly requested during day $d$. If the signal is energy constrained, only a fraction of the worst case reserve energy $\int_{t=0}^{N-1} r(t) dt$ will be requested as consumption decrease from the buildings. The rest part of $\int_{t=0}^{N-1} r(t) dt$ will be stored as thermal energy in the buildings and will reduce the required heating/cooling energy (and the respective costs) during day $d+1$. For this reason, the buildings are willing to provide reserves even if $k/c<1$. We present simulation results for PEC in winter and summer in Fig.~\ref{fig:CapacityPayment}b. The threshold ratio for reserve provision is \mbox{$k/c=0.32$} for both winter and summer, and it depends on $\varepsilon$ and $T$. The capacity increases slowly in the ratio range $[0.32-0.99]$, particularly in summer, and then it suddenly jumps to higher values as $k/c$ approaches to $1$. No results are shown for PC in Fig.~\ref{fig:CapacityPayment}b because the capacity is zero for $k/c<1$, as explained before.

Fig.~\ref{fig:CapacityPayment}a shows that for unconstrained SFC signals (PC) a ratio $k/c=1.01$, which means a capacity payment $1\%$ higher than the electricity price, taps most of the reserve potential. In particular, further increasing the ratio up to the maximum considered value $k/c=2$ increases the reserves only by $8\%$ in winter and no more than $18\%$ in summer. Assuming an average electricity price of $146.6$~CHF/MWh, which is the case for consumers who consume more than $60$~MWh/year in Zurich, with a ratio $k/c=1.01$ capacity payments around $148$~CHF/MW/h are needed. This is significantly lower than the most expensive accepted bids, but approximately $4$~times higher than the average capacity payment in $2013$ \cite{swissgridOnline}. Fig.~\ref{fig:CapacityPayment}b shows that energy-constrained SFC signals can reduce the necessary capacity payments down to $32\%$ of the retail price ($k/c=0.32$), but of course with a large reduction in the reserve capacity. Thus, reserves are actually costly for buildings already equipped with MPC for \emph{energy efficient} (optimal) control, especially if the SFC signal is not energy-constrained. This is in contrast to TCLs with simple hysteresis control based on a deadband, where reserves can be provided at a lower cost. However, note that our calculations are based on the prevailing case where the buildings acquire energy in the retail electricity market. In another market setting where the buildings acquired energy directly in the spot market, the buildings could offer reserves at more competitive prices because the retail electricity prices are typically significantly higher than the wholesale spot electricity prices. The analysis of this section provides intuition on the relationship between the amount of reserves from building aggregations and the capacity payments. In practice, estimating the capacity payment is a challenging task that needs to consider additional costs, e.g., due to device wear, but also the competition, i.e., the expected bid prices of generators and/or other load aggregations.

\subsection{Reserve Product Characteristics} \label{ResProdChar}
In this section, we fix $k/c=1.1$ and investigate the influence of important reserve product characteristics on reserve capacities. For PC, we consider reserves with daily or hourly duration, and symmetric or asymmetric capacities. For PEC, we investigate reserves with daily or hourly duration, as well as different ($T$, $\varepsilon$) pairs, based on Table~\ref{tab:LFCanalysis}. We summarize simulation results with respect to the total reserve capacity for two weeks in winter and summer in Tables~\ref{tab:CapacityAndReserveCharacteristics1} and \ref{tab:CapacityAndReserveCharacteristics2}, where positive values denote up- and negative values down-reserves.

For PC and symmetric products\footnote{Note that the results of Table~\ref{tab:CapacityAndReserveCharacteristics1} are not directly comparable with the values reported in Tables 1 and 2 of \cite{VrettosIFAC2014}, because different COP values have been used in the simulations of the two papers.}, hourly reserves increase capacities by $2.9\%$ in winter and $3.3\%$ in summer, compared to daily reserves. If asymmetric reserves are allowed, the aggregation provides significantly more down- than up-reserves in summer, whereas in winter no up-reserves are provided at all. Down-reserves (increasing power consumption) are preferable for buildings equipped with MPC for energy efficient control because the capacity can be offered without increased baseline consumption. The energy efficient control tries to stay as close as possible to the minimum power consumption of the heating/cooling device. In order to provide up-reserves (consumption decrease), a building needs to be able to reduce its consumption without violating occupant comfort. Therefore, the building must schedule its operation (baseline) at a power level higher than the energy optimal, which increases energy costs. On the other hand, down-reserves can be provided while operating at the energy efficient trajectory because the consumption can only increase will tracking the SFC signal.

Compared to PC, PEC increase the reserves for all pairs ($T$, $\varepsilon$) up to $3.8\%$ in winter and $5.6\%$ in summer. For given $T$ and $\varepsilon$, adopting hourly instead of daily reserves increases the capacities up to $2\%$ in winter and $9.6\%$ in summer. Note that the increase is higher for large $T$. Notice that shorter $T$ are characterized by smaller $\varepsilon T$ products, and so constraint \eqref{EnergyCons} on the uncertainty becomes tighter. Therefore, one would intuitively expect that decreasing $T$ increases the reserve capacities monotonically. However, one has to consider that longer $T$ couple more optimization periods, which is why a monotonic behavior is not observed in our simulation results. To gain deeper insight into the dependence of reserve capacities on $T$ and $\varepsilon$, we run simulations for two weeks in winter and summer with $T \in \{1, 2, 4, 6, 8, 12\}$~hours and $\varepsilon$ varying from $0.05$ to $0.5$ with $0.05$ increments, i.e., $60$ combinations for each season. We assume daily, symmetric reserves and simulate only Lv$1$, i.e., the reserve scheduling problem, whereas the building HVAC control is not considered. The reason is that many of the combinations of $T$ and $\varepsilon$ are not achievable by the Chebyshev filter \eqref{Chebyshev}, and so Lv$2$ will likely be infeasible. However, our analysis provides intuition on the effect of $T$ and $\varepsilon$ that could be useful for filter design in a practical application.

\begin{table}[t]
\renewcommand{\arraystretch}{1.09}
\caption{Capacity (MW) of Reserve Products (Power Constraints)}
\centering
\begin{tabular}{cc|cccc}
\hline
Duration   & Symmetry  &  \multicolumn{2}{c}{Winter}    &  \multicolumn{2}{c}{Summer} \\
\hline
daily  &  symmetric &  \multicolumn{2}{c}{$\pm 105.9$}     &  \multicolumn{2}{c}{$\pm 89.6$}  \\
daily  &  asymmetric  &  \multicolumn{2}{c}{$+ 0 / -196.9$}   &  \multicolumn{2}{c}{$+45.3 / -195.0$}  \\
hourly  &  symmetric  &  \multicolumn{2}{c}{$\pm 109.0$} &  \multicolumn{2}{c}{$\pm 92.6$}  \\
hourly  &  asymmetric  &  \multicolumn{2}{c}{$+0 / -213.6$}   &  \multicolumn{2}{c}{$+44.8 / -198.4$}  \\
\hline
\end{tabular}
\label{tab:CapacityAndReserveCharacteristics1}
\end{table}

\begin{table}[t]
\renewcommand{\arraystretch}{1.09}
\caption{Capacity (MW) of Reserve Products (Power \& Energy Constr.)}
\centering
\begin{tabular}{ccc|cccc}
\hline
$T$ (h) & $\varepsilon$ (-) & $\varepsilon T$ (h) & day/win & hour/win & day/sum & hour/sum\\
\hline
$1$ & $0.382$ & $0.382$ & $\pm 109.4$ & $\pm 110.0$ & $\pm 92.5$ & $\pm 97.6$\\
$2$ & $0.300$ & $0.600$ & $\pm 109.9$ & $\pm 110.7$ & $\pm 93.1$ & $\pm 99.3$\\
$4$ & $0.317$ & $1.268$ & $\pm 109.0$ & $\pm 110.2$ & $\pm 92.6$ & $\pm 99.5$\\
$6$ & $0.290$ & $1.740$ & $\pm 108.9$ & $\pm 110.4$ & $\pm 93.3$ & $\pm 99.6$\\
$8$ & $0.237$ & $1.896$ & $\pm 108.9$ & $\pm 111.0$ & $\pm 94.6$ & $\pm 101.5$\\
$12$ & $0.203$ & $2.436$ & $\pm 107.7$ & $\pm 111.1$ & $\pm 92.8$ & $\pm 99.6$\\
\hline
\end{tabular}
\label{tab:CapacityAndReserveCharacteristics2}
\end{table}

\begin{figure}[t]%
\centering
\begin{minipage}{0.25\linewidth}%
\centering \includegraphics[width=2.45in,height=1.8in]{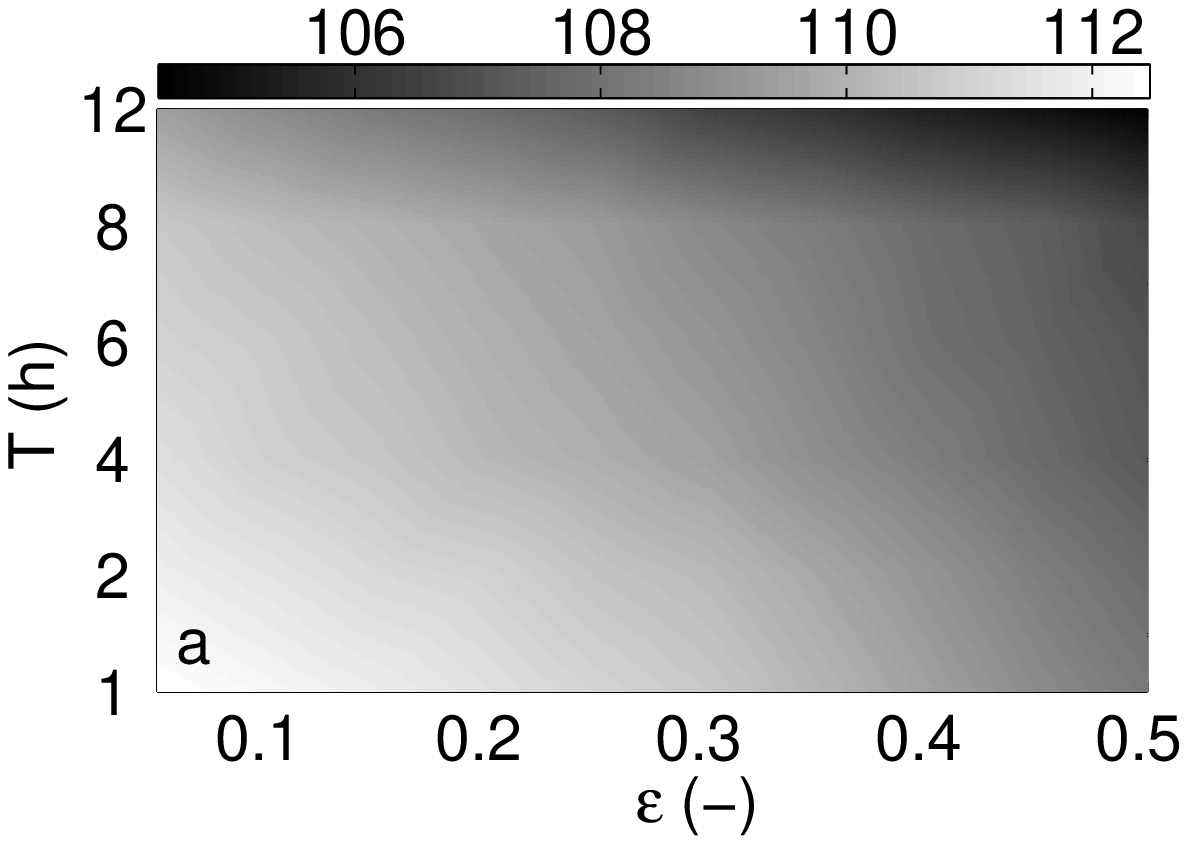}
\end{minipage}%
\begin{minipage}{0.94\linewidth}%
\centering \includegraphics[width=2.45in,height=1.8in]{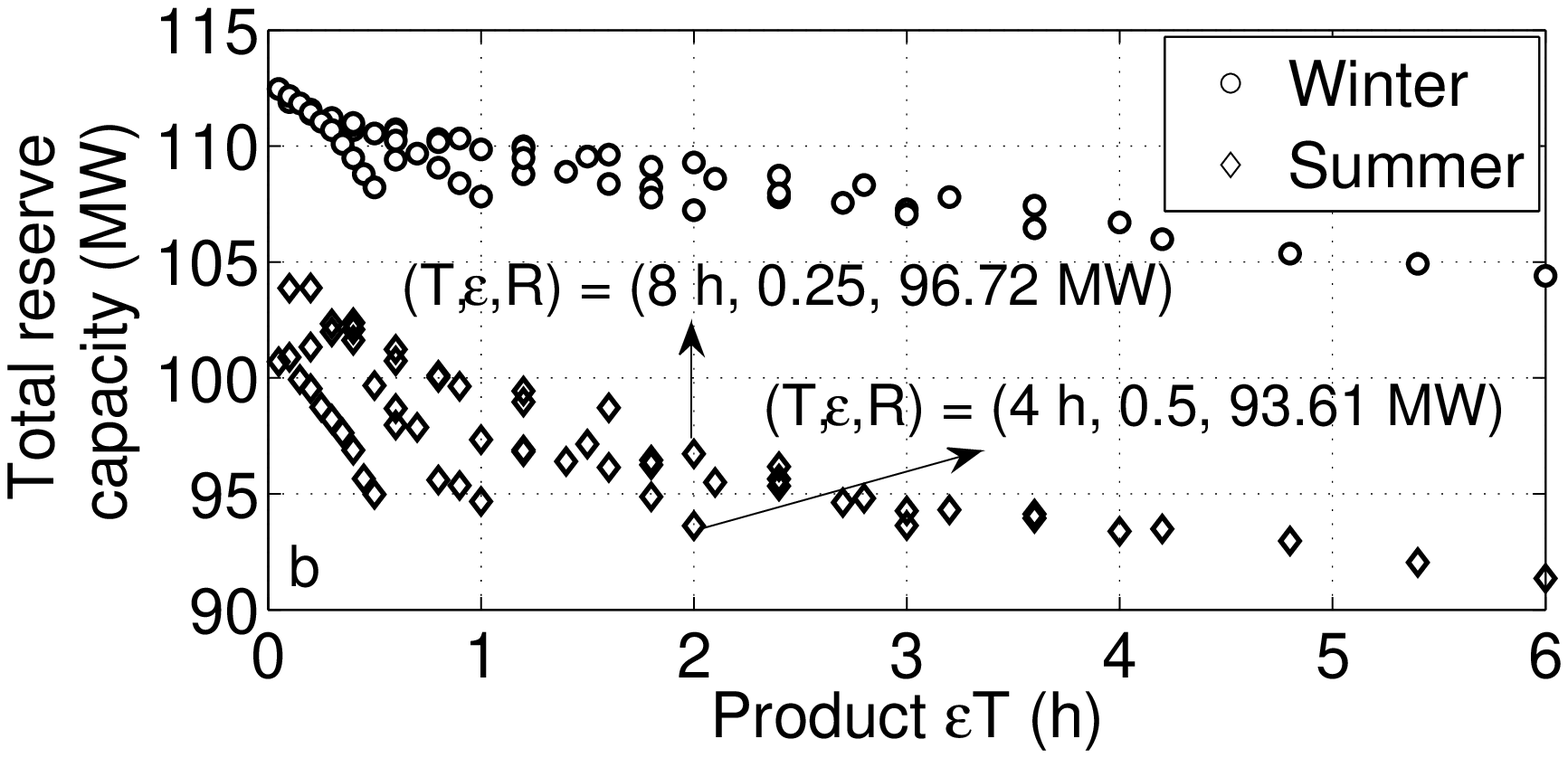}
\end{minipage}%
\caption{Left: sensitivity of reserve capacity (MW) on averaging period $T$ and bias $\varepsilon$ in winter. Right: capacity's dependence on $\varepsilon T$ in winter and summer.}%
\label{fig:TaEpsAnalysis}%
\end{figure}

In Fig.~\ref{fig:TaEpsAnalysis}(a), we show the total reserve capacities for each of the simulated cases in winter. Similar results are obtained for summer, but are omitted here due to space limitations. The dependence of reserve capacities on $T$ and $\varepsilon$ demonstrates a clear pattern: increasing any of the two parameters reduces the reserve capacity. Note that the capacity is more sensitive to $\varepsilon$ than to $T$. To better illustrate this, we present the results of Fig.~\ref{fig:TaEpsAnalysis}(a) based on the product $\varepsilon T$ in Fig.~\ref{fig:TaEpsAnalysis}(b). As expected, increasing $\varepsilon T$ generally decreases the reserve capacity. The same $\varepsilon T$ can be obtained by different ($T$, $\varepsilon$) pairs: for example, both $(8~\textrm{h}, 0.25)$ and $(4~\textrm{h}, 0.5)$ obtain $\varepsilon T=2$, but the first achieves a capacity $96.72$~MW, whereas the second achieves $93.61$~MW. In this case, the capacity is more sensitive to $\varepsilon$ than to $T$. We compared all cases in winter and summer with the same $\varepsilon T$ and found out that in $70\%$ of them smaller $\varepsilon$ is preferable to smaller $T$. This means that buildings can cope easier with signals that are significantly energy constrained over long periods than signals that are moderately constrained over shorter periods.

\section{Concluding Remarks} \label{Conclusion}
In this paper, we proposed a new framework based on robust optimization and MPC for scheduling and provision of secondary frequency control (SFC) reserves by the HVAC systems of commercial building \emph{aggregations}. The framework incorporates tractable methods to account for energy-constrained SFC signals, and relies on decentralized reserve provision to keep real-time communication requirements low and preserve privacy. We also presented how the framework can be used to estimate the SFC reserve potential from commercial buildings.

Our analysis was based on four main assumptions: (a) there is no plant-model mismatch; (b) the predictions of weather and occupancy are perfect; (c) all building states can be measured; and (d) the reaction of heating/cooling devices is fast and does not cause any wear. Therefore, the reported results provide an upper bound on the amount of reserves from buildings. For a real implementation, additional care must be taken for (a)-(d): accounting for modeling and weather/occupancy prediction errors, use of state estimators, and modeling of the fast dynamics of heating/cooling devices. If multi-zone building models with hundreds of states are available, model reduction techniques \cite{Sturzenegger2013Experiment} can be used to reduce the optimization problem's size. The reserve scheduling problem considered capacity payments, but neglected revenues from reserve energy utilization. Incorporating the latter in the framework is possible if SFC signal scenarios can be generated based on historical data, and is expected to reduce the necessary capacity payments. Additionally, comfort constraints can be relaxed as chance constraints allowing comfort zone violations with a small probability, which is typical for building climate control, while keeping HVAC input constraints robust. Such combination of robust and chance-constrained optimization is likely to increase reserve capacities. In the future, we plan to include these aspects in the framework, implement and test it on a real building.

Overall, our results show that significant amounts of SFC reserves can be \emph{reliably} offered by an aggregation of {\raise.17ex\hbox{$\scriptstyle\sim$}}$100$ commercial buildings without loss of occupant comfort. We found that with traditional unconstrained SFC signals asymmetric reserves are preferable for buildings, that energy-constrained SFC signals reduce the necessary capacity payments and increase reserves by up to {\raise.17ex\hbox{$\scriptstyle\sim$}}$10\%$ compared to traditional SFC signals, and that reducing the duration of reserve products from $1$ day to $1$ hour increases reserves by up to {\raise.17ex\hbox{$\scriptstyle\sim$}}$9\%$.

\bibliographystyle{elsarticle-num}
\bibliography{biblio_EV}

\end{document}